\documentclass{theoretics}

\usepackage{bbm}
\usepackage[capitalize,nameinlink,noabbrev]{cleveref}

\usepackage{nicefrac}
\renewcommand{\epsilon}{\varepsilon}

\providecommand{\ssum}{\sqcup}
\DeclareMathOperator{\round}{round}
\DeclareMathOperator{\dist}{dist}
\DeclareMathOperator{\supp}{supp}
\DeclareMathOperator*{\EE}{\mathbb{E}}
\providecommand{\bbone}{\mathbbm{1}}
\providecommand{\indicator}{\mathbbm{1}}
\providecommand{\mexp}{M}
\providecommand{\nexp}{\nu}
\providecommand{\expT}{R}

\title{Sparse juntas on the biased hypercube}
\ThCSauthor[weizmann]{Irit Dinur}{irit.dinur@weizmann.ac.il}[0000-0002-4335-5237]
\ThCSaffil[weizmann]{Weizmann Institute of Science, Rehovot, Israel}
\ThCSauthor[technion]{Yuval Filmus}{yuvalfi@cs.technion.ac.il}[0000-0002-1739-0872]
\ThCSaffil[technion]{Technion --- Israel Institute of Technology, Haifa, Israel}
\ThCSauthor[tata]{Prahladh Harsha}{prahladh@tifr.res.in}[0000-0002-2739-5642]
\ThCSaffil[tata]{Tata Institute of Fundamental Research, Mumbai, India}

\ThCSthanks{An extended abstract of results from the earlier version of this paper and the hypergraph agreement theorem \cite{DinurFH-agree} appeared in SODA 2019 \cite{DinurFH2019}. Irit Dinur was supported by ERC grant 772839, and ISF grant 2073/21. Yuval Filmus was supported by the European Union's Horizon 2020 research and innovation programme under grant agreement No~802020-ERC-HARMONIC.  Prahladh Harsha's research was supported by the Department of Atomic Energy, Government of India, under project 12-R\&D-TFR-5.01-0500 and in part by UGC-ISF grant and Swarnajayanti Fellowship. Part of the work was done when Prahladh Harsha was visiting the Weizmann Institute of Science.}

\ThCSshortnames{I. Dinur, Y. Filmus, P. Harsha}
\ThCSshorttitle{Sparse juntas on the biased hypercube}

\ThCSkeywords{Boolean functions, $p$-biased distribution, Junta theorem, sparse functions}

\ThCSyear{2024}
\ThCSarticlenum{18}
\ThCSreceived{Jun 30, 2023}
\ThCSrevised{Mar 26, 2024}
\ThCSaccepted{May 25, 2024}
\ThCSpublished{Jul 30, 2024}
\ThCSdoicreatedtrue

\begin{document}

\maketitle

\begin{abstract}
We give a structure theorem for Boolean functions on the $p$-biased hypercube which are $\epsilon$-close to degree~$d$ in $L_2$, showing that they are close to \emph{sparse juntas}.
Our structure theorem implies that such functions are $O(\epsilon^{C_d} + p)$-close to constant functions. We pinpoint the exact value of the constant $C_d$.

We also give an analogous result for monotone Boolean functions on the biased hypercube which are $\epsilon$-close to degree~$d$ in $L_2$, showing that they are close to \emph{sparse DNFs}.

Our structure theorems are optimal in the following sense: for every $d,\epsilon,p$, we identify a class $\mathcal{F}_{d,\epsilon,p}$ of degree~$d$ sparse juntas which are $O(\epsilon)$-close to Boolean (in the monotone case, width~$d$ sparse DNFs) such that a Boolean function on the $p$-biased hypercube is $O(\epsilon)$-close to degree~$d$ in $L_2$ if and only if it is $O(\epsilon)$-close to a function in $\mathcal{F}_{d,\epsilon,p}$.
\end{abstract}

\section{Introduction}
\label{sec:introduction}

Let $f\colon \{0,1\}^n \to \{0,1\}$ be a Boolean function on the hypercube, where $\{0,1\}^n$ is endowed with the uniform measure. If $f$ has degree $1$ then it is a dictator, and if it has degree $d$ then it is a junta~\cite{NisanS1994}. Friedgut, Kalai, and Naor (FKN) ~\cite{FriedgutKN2002} showed that if $f$ is $\epsilon$-close to degree~$1$ (that is, $\|f^{>1}\|^2 \leq \epsilon$) then $f$ is $O(\epsilon)$-close to a dictator (that is, $\Pr[f \neq g] = O(\epsilon)$ for some dictator $g$). Kindler and Safra~\cite{KindlerS2002,Kindler2003} (see also \cite{KellerK2020}) showed that if $f$ is $\epsilon$-close to degree~$d$ then $f$ is $O(\epsilon)$-close to a $2^{O(d)}$-junta.

Kindler and Safra proved their result in the more general setting of the biased hypercube, which is $\{0,1\}^n$ endowed with the biased measure $\mu_p$. They showed that if $f$ is a Boolean function on the $p$-biased hypercube which is $\epsilon$-close to degree $d$ then $f$ is $O(\epsilon + \epsilon^{5/4}/p^{4d})$-close to an $O(1/p^{4d})$-junta, assuming $p \leq 1/2$. When $p$ is constant, this gives the same statement as before, but when $p$ is small, the result is no longer tight.

The biased hypercube in the small $p$ regime is interesting for at least two reasons. First, it is the regime that is most relevant for understanding threshold phenomena in the Erd\H{o}s--R\'enyi random graph model $G(n,p)$. Second, it is a good model for other domains with sparse inputs, such as the symmetric group and the Grassmann scheme, the latter of which was recently used to prove the 2-to-2 games conjecture~\cite{KhotMS2023}.

In this paper, we extend the FKN and Kindler--Safra theorems to functions on the biased hypercube, proving a statement which does not deteriorate as $p$ gets small. For every $d,\epsilon$ and $p \leq 1/2$ we find an ``explicit" set of Boolean functions $\mathcal{F}_{d,\epsilon,p}$ (consisting of degree $d$ ``sparse juntas'', a concept we define below) such that
\begin{enumerate}[(a)]
\item If $f\colon (\{0,1\}^n,\mu_p) \to \{0,1\}$ is $\epsilon$-close to degree~$d$ then $f$ is $O(\epsilon)$-close to a function in $\mathcal{F}_{d,\epsilon,p}$.
\item Every function in $\mathcal{F}_{d,\epsilon,p}$ is $O(\epsilon)$-close to degree $d$ with respect to $\mu_p$.
\end{enumerate}

Let us demonstrate what we mean by \emph{explicit} using the case $d = 1$, which appears in earlier work of Filmus~\cite{Filmus2016-fkn}. Filmus showed that we can take
\[
 \mathcal{F}_{1,\epsilon,p} =
 \left\{
 \bigvee_{i \in S} y_i,
 \lnot \bigvee_{i \in S} y_i :
 |S| \leq \max(1, C\sqrt{\epsilon}/p)
 \right\},
\]
for some constant $C$, where $(y_1,\ldots,y_n) \in \{0,1\}^n$ is the input point.

The functions in $\mathcal{F}_{1,\epsilon,p}$ are $O(\sqrt{\epsilon})$-close to dictators and so $O(\sqrt{\epsilon} + p)$-close to some constant function, and the same holds for every Boolean function which is $\epsilon$-close to degree~$1$. This shows that if we consider Boolean functions which are $\epsilon$-close to degree~$1$ up to an error of magnitude $\sqrt{\epsilon}$, then all we see is dictators. If we look closer, narrowing the magnitude of the error to $\epsilon$, then a more nuanced picture emerges. We will be interested in both points of view. We state our results formally in \Cref{sec:intro-main-results}.

When $d > 1$, one cannot hope for a set $\mathcal{F}_{d,\epsilon,p}$ which is as explicit as in the case $d = 1$. Indeed, even a complete description of all Boolean degree~$d$ functions (corresponding to the case $\epsilon = 0$) is not available. The set $\mathcal{F}_{d,\epsilon,p}$ that we give is as explicit as we were able to make it; perhaps the reader can improve it. Moreover, it allows us to show that Boolean functions which are $\epsilon$-close to degree~$d$ are $O(\epsilon^C)$-close to Boolean degree~$d$ functions, for an appropriate constant $C$, demonstrating the applicability of our main structure theorem.

\medskip

The proof of Kindler and Safra deteriorates with $p$ due to the use of hypercontractivity. In contrast, our proof works by reduction to the unbiased setting, and so does not suffer from this deterioration. In more detail, we express the biased hypercube as a weighted average of unbiased hypercubes of smaller dimension:
\[
 (\{0,1\}^n, \mu_p) \approx (\{0,1\}^{\mathbf{S}}, \mu_{1/2}), \text{ where } \mathbf{S} \sim \mu_{2p}.
\]
We apply the \emph{unbiased} Kindler--Safra theorem on each unbiased hypercube $\{0,1\}^S$, and combine the results using an \emph{agreement theorem}. We describe our methods in more detail in \Cref{sec:intro-proof-sketch}.

\subsection{Main results}
\label{sec:intro-main-results}

Here is the generalization of the result of Filmus~\cite{Filmus2016-fkn} to arbitrary $d$. The statement uses various terms which we explain after it. The statement is heavily inspired by the FKN theorem for functions on the symmetric group due to Filmus~\cite{Filmus2021}.

\begin{theorem}[Main] \label{thm:main-intro}
Suppose that $f\colon (\{0,1\}^n, \mu_p) \to \{0,1\}$ is $\epsilon$-close to degree~$d$, where $p \leq 1/2$. Then $f$ is $O(\epsilon)$-close to $\round(g, \{0,1\})$, where $g$ is a degree~$d$ polynomial satisfying the following properties, for some constant $C$ depending only on $d$:
\begin{enumerate}[(i)]
\item If $y \in \{0,1\}^n$ has (Hamming) weight at most~$d$ then $g(y) \in \{0,1\}$.
\label{itm:main-boolean}
\item For every $T$ and $e$, the monomial support of $g$ contains at most $C/p^e$ sets $U \supseteq T$ of size $|T| + e$.
\label{itm:main-sparse}
\item For all $e$, the number of inputs $y \in \{0,1\}^n$ of (Hamming) weight $e$ such that $g(y) \notin \{0,1\}$ but $g(z) \in \{0,1\}$ for all $z < y$ is at most $C\epsilon/p^e$.
\label{itm:main-non-boolean}
\end{enumerate}

Conversely, if $g$ is a degree~$d$ polynomial satisfying these properties then $\round(g, \{0,1\})$ is $O(\epsilon)$-close to degree~$d$.
\end{theorem}

Rather than specifying the list $\mathcal{F}_{d,\epsilon,p}$ directly, we instead specify a list $\mathcal{G}_{d,\epsilon,p}$ of degree~$d$ functions. The functions in $\mathcal{F}_{d,\epsilon,p}$ are obtained by rounding the functions in $\mathcal{G}_{d,\epsilon,p}$ to Boolean. The corresponding list for $d = 1$ is
\[
 \mathcal{G}_{1,\epsilon,p} =
 \left\{
 \sum_{i \in S} y_i,
 1 - \sum_{i \in S} y_i :
 |S| \leq \max(1, C\sqrt{\epsilon}/p)
 \right\}.
\]

A degree~$d$ polynomial $g$ is in $\mathcal{G}_{d,\epsilon,p}$ if it satisfies the three properties listed in the theorem. To explain these properties, let us introduce the \emph{monomial expansion} of $g$, which is its unique expansion as a linear combination of monomials $y_S = \prod_{i \in S} y_i$. We denote the coefficient of $y_S$ by $\tilde{g}(S)$; these are the \emph{monomial coefficients} of $g$.

The monomial expansion differs from the more usual Fourier expansion, in which instead of the monomials $y_S$ we have the Fourier characters $\omega_S = \prod_{i \in S} \frac{y_i - p}{\sqrt{p(1-p)}}$. Using the monomial expansion guarantees that the functions in $\mathcal{G}_{d,\epsilon,p}$ have the same form regardless of $p$ (cf.\ $\sum_{i \in S} y_i$ to $\sum_{i \in S} \sqrt{p(1-p)} \frac{y_i - p}{\sqrt{p(1-p)}} + p|S|$). Concretely, the monomial coefficients appearing in the functions in $\mathcal{G}_{d,\epsilon,p}$ depend only on $d$, a property which fails for the Fourier expansion. 

\textbf{\Cref{itm:main-boolean}}
states that $g(y) \in \{0,1\}$ whenever $y$ has (Hamming) weight at most~$d$. Equivalently, if $|S| \leq d$ then
\[
 \sum_{T \subseteq S} \tilde{g}(T) \in \{0, 1\}.
\]
This implies that the monomial coefficients of $g$ are bounded integers (implying that $g$ is integer-valued, as a function on $\{0,1\}^n$). As an example, when $d = 1$, the constant coefficient is either $0$ or $1$; in the former case, all other coefficients are $0$ or $1$; and in the latter case, all other coefficients are $0$ or $-1$.

\textbf{\Cref{itm:main-sparse}}
concerns the \emph{monomial support} of $g$, which is the set
\[
 \supp(g) = \{ S \subseteq [n] : \tilde{g}(S) \neq 0 \}.
\]
Choosing $T = \emptyset$, the property states that the expected number of monomials in $\supp(g)$ which evaluate to~$1$ under a random input is $O(1)$. Considering other $T$, the same holds even under the condition $y_T = 1$. For these reasons, we call a polynomial satisfying this property a \emph{sparse junta}.

This item fails for the Fourier expansion. For example, the Fourier support of the function $\sum_{i=1}^{C/p^2} y_{2i-1} y_{2i}$ contains $2C/p^2$ degree-1 monomials rather than just $O(1/p)$.

\textbf{\Cref{itm:main-non-boolean}}
concerns \emph{minimal non-Boolean inputs}. An input $y \in \{0,1\}^n$ is a minimal non-Boolean input of $g$ if $g(y) \notin \{0,1\}$ but $g(z) \in \{0,1\}$ for all inputs $z$ strictly below $y$, that is, obtained from $y$ by switching one or more $1$s to $0$s. 
Equivalently, if $y = \bbone_S$ then the restriction $g|_S$ obtained by zeroing out all coordinates outside of $S$ is Boolean except for the input $\bbone_S$. Due to property \Cref{itm:main-boolean}, every minimal non-Boolean input occurring in $g$ has (Hamming) weight at least $d+1$.

We can think of a minimal non-Boolean input as a forbidden configuration, which we can describe by specifying the function $g|_S$. When $d = 1$, there are two minimal non-Boolean inputs, corresponding to the functions
\[
 y_1 + y_2 \text{ and } 1 - y_1 - y_2.
\]
If $\tilde{g}(\emptyset) = 0$ then only the first one $y_1 + y_2$ is possible. Defining $m = |\supp(g)|$, the number of minimal non-Boolean inputs is $\binom{m}{2}$, and so \Cref{itm:main-non-boolean} states that $\binom{m}{2} = O(\epsilon/p^2)$. This implies that either $m \leq 1$ or $m = O(\sqrt{\epsilon}/p)$. The case $\tilde{g}(\emptyset) = 1$ is similar.

We list all minimal non-Boolean inputs for $d = 2$ in \Cref{sec:forbidden-configurations}, where we also show that the number of such inputs in independent of $n$.

\subsubsection*{Monotone version}

When the function $f$ is monotone, we can guarantee that the function $g$ is monotone as well, in fact a \emph{monotone width~$d$ DNF}, which is a disjunction of \emph{minterms} $y_{i_1} \land \cdots \land y_{i_e}$ with $e \leq d$.

\begin{theorem} \label{thm:main-monotone-intro}
Suppose that $f\colon (\{0,1\}^n, \mu_p) \to \{0,1\}$ is a monotone function which is $\epsilon$-close to degree~$d$, where $p \leq 1/2$. Then $f$ is $O(\epsilon)$-close to a monotone width~$d$ DNF $g$, where $g$ satisfies the following properties, for some constant $C$ depending only on $g$:
\begin{enumerate}[(i)]
\item For every $T$ and $e$, the DNF $g$ contains at most $C/p^e$ minterms of $U \supseteq T$ of size $|U| + e$.
\label{itm:main-monotone-sparse}
\item For all $e$, the number of sets $S$ of size $e$ such that $\deg(g|_S) > d$ but $\deg(g|_T) \leq d$ for all $T \subsetneq S$ is at most $C\epsilon/p^e$.
\label{itm:main-monotone-high}
\end{enumerate}
Conversely, if $g$ is a monotone width $d$ DNF satisfying these properties then it is $O(\epsilon)$-close to degree~$d$.
\end{theorem}

When $d = 1$, the unique forbidden configuration (corresponding to \textbf{\Cref{itm:main-monotone-high}}) is $y_1 \lor y_2$. We list all forbidden configurations for $d = 2$ in \Cref{sec:forbidden-configurations}, where we also show that there is a finite number of them for every~$d$.

\subsubsection*{Junta approximation}

The FKN theorem of Filmus~\cite{Filmus2016-fkn} implies that if $f\colon (\{0,1\}^n, \mu_p) \to \{0,1\}$ is $\epsilon$-close to degree~$1$ then it is $O(\sqrt{\epsilon})$-close to a dictator and $O(\sqrt{\epsilon}+p)$-close to a constant. \Cref{thm:main-intro} implies a similar result for arbitrary~$d$, where $\sqrt{\epsilon}$ is replaced by an appropriate power of $\epsilon$.

\begin{theorem}[Junta approximation] \label{thm:junta-intro}
Suppose that $f\colon (\{0,1\}^n,\mu_p) \to \{0,1\}$ is $\epsilon$-close to degree~$d$, where $p \le 1/2$. There are parameters $m,\mexp$, depending only on $d$, such that:
\begin{enumerate}[(a)]
\item $f$ is $O(\epsilon^{1/\mexp})$-close to a Boolean degree $d$ function.
\item $f$ is $O(\epsilon^{1/m}+p)$-close to a Boolean constant function.
\end{enumerate}
 \end{theorem}

 When $d = 1$, the optimal value of both $m$ and $\mexp$ is~$2$, which is tight when $f$ is obtained by rounding the following function to Boolean:
 \[
  g = y_1 + \cdots + y_{\sqrt{\epsilon}/p}.
 \]
 
 We can generalize this construction to arbitrary $d$. Let $P$ be a degree~$d$ polynomial such that $P(0),\ldots,\allowbreak {P(m-1)} \in \{0,1\}$ and $P(0) \neq P(1)$ (von zur Gathen and Roche~\cite{GathenR1997} studied such non-constant polynomials without the restriction $P(0) \neq P(1)$), and take the degree~$d$ polynomial
 \[
  g = P(y_1 + \cdots + y_{\epsilon^{1/m}/p}).
 \]

The input to $P$ behaves roughly like a Poisson random variable with expectation $\lambda = \epsilon^{1/m}$, and so $\Pr[g \notin \{0,1\}] = O(\lambda^m) = O(\epsilon)$. This implies that $\round(g, \{0,1\})$ is $O(\epsilon)$-close to degree~$d$.

Since $P(1) \neq P(0)$, a similar calculation shows that $\Pr[g = P(0)] \approx 1 - \Theta(\lambda)$ and that $\Pr[g = P(1)] \approx \Theta(\lambda)$, and so $g$ is $\Theta(\lambda)$-far from all constant functions.

Using a Ramsey-theoretic argument, we show that this kind of construction is the best possible for $m$.

\begin{theorem} \label{thm:junta-intro-formula}
Given $d$, the optimal value of $m$ in \Cref{thm:main-intro} is the maximum $m$ such that there exists a degree~$d$ polynomial $P$ satisfying $P(0),\ldots,P(m-1) \in \{0,1\}$ and $P(0) \neq P(1)$.
\end{theorem}

As an example, when $d = 2$ we get $m = 4$, corresponding to the polynomial $P(z) = \frac{z(3-z)}{2}$.

It is an intriguing open question to find the optimal value of $\mexp$. Another interesting exponent is the best $\nexp$ such that $f$ is $O(\epsilon^{1/\nexp})$-close to some Boolean junta or to some degree~$d$ junta (both exponents are the same).

\subsection{Proof sketch}
\label{sec:intro-proof-sketch}

We prove \Cref{thm:main-intro} by reduction to the Kindler--Safra theorem, in the following form.

\begin{theorem}[Kindler--Safra] \label{thm:kindler-safra-intro}
If $f\colon (\{0,1\}^n, \mu_{1/2}) \to \{0,1\}$ is $\epsilon$-close to degree~$d$ then there is a Boolean degree~$d$ function $g$ which is $O(\epsilon)$-close to $f$.
\end{theorem}

\noindent (While Kindler and Safra did not state their theorem in this way, this formulation easily follows from their result.)

Suppose we are given a function $f\colon (\{0,1\}^n, \mu_p) \to \{0,1\}$ which is $\epsilon$-close to degree~$d$. We look at restrictions $f|_S$ obtained by zeroing out the coordinates outside of $S$. If $S \sim \mu_{2p}$ then on average, $f|_S$ is close to degree~$d$ with respect to the uniform measure on $\{0,1\}^S$:
\[
 \EE_{S \sim \mu_{2p}}\left[ \EE_{\mu_{1/2}}[(f|_S - f^{\leq d}|_S)^2] \right] = \EE_{\mu_p}[(f - f^{\leq d})^2] \leq \epsilon.
\]
Indeed, if $S \sim \mu_{2p}$ and $y \sim \mu_{1/2}(\{0,1\}^S)$ then the marginal distribution of $y$ is $\mu_p$.

Applying the Kindler--Safra theorem for every $f|_S$, we obtain a Boolean degree~$d$ function $g_S$ such that
\begin{equation} \label{eq:proof-sketch-1}
 \EE_{S \sim \mu_{2p}}\left[ \EE_{\mu_{1/2}}[(f|_S - g_S)^2] \right] = O(\epsilon).
\end{equation}

We would like to construct the function $g$ by pasting together the various functions $g_S$, and for that we need to know that the $g_S$ agree with each other, on average. This sort of pasting is achieved using \emph{agreement theorems}, and in this case we use the following theorem, with $K$ being the maximal size of the monomial support of a Boolean degree~$d$ function, $q = 2p$, and $c = \sqrt{1/2}$.

\begin{theorem}[Junta agreement theorem] \label{thm:agreement-intro}
Fix the following parameters: integer $K$ (junta size), $q \in (0,1)$ (bias), $c \in (0,1)$ (fractional intersection size).

For each $S \subseteq [n]$, let $g_S$ be a degree~$d$ polynomial whose monomial support contains at most~$K$ monomials.

Suppose that the following agreement condition holds:
\[
 \Pr_{\substack{T \sim \mu_{cq} \\ T_1,T_2 \sim \mu_r}}[g_{T \cup T_1}|_T \neq g_{T \cup T_2}|_T] \leq \epsilon, \text{ where } r = \frac{(1-c)q}{1-cq}.
\]
(The value of $r$ is chosen so that the marginal distribution of $S_1 := T \cup T_1$ and $S_2 := T \cup T_2$ is $\mu_q$.)

Then there exists a degree~$d$ polynomial $g$ such that
\[
 \Pr_{S \sim \mu_q}[g|_S \neq g_S] = O_{K,c}(\epsilon).
\]
(The error bound depends on $K$ and $c$ but is independent of $q$.)

Moreover, for each $A$, the monomial coefficient $\tilde{g}(A)$ is chosen by majority decoding: it maximizes
\[
 \Pr_{\substack{S \sim \mu_{2p} \\ S \supseteq A}}[\tilde{g}_S(A) = \tilde{g}(A)].
\]
\end{theorem}

The junta agreement theorem follows from the more general agreement theorem in our earlier work~\cite{DinurFH-agree}, which does not require a bound on the size of the monomial support. However, since the proof of the junta agreement theorem is much easier than the proof of the general agreement theorem, we include it in this paper.

In order to apply \Cref{thm:agreement-intro}, we need to show that the condition in the theorem holds.  Since $f|_{S_1\cap T} = f|_{S_2 \cap T}$, \Cref{eq:proof-sketch-1} implies that
\[
 \EE_{S_1,S_2} [(g_{S_1}|_T - g_{S_2}|_T)^2] = O(\epsilon).
\]
Since $g_{S_1},g_{S_2}$ are both juntas, either $g_{S_1}|_T = g_{S_2}|_T$ or $\Pr[g_{S_1}|_T \neq g_{S_2}|_T] = \Omega(1)$, and this implies the agreement condition of \Cref{thm:agreement-intro}.

Applying \Cref{thm:agreement-intro}, we obtain a function $g$ such that
\[
 \Pr_{S \sim \mu_p}[g|_S \neq g_S] = O(\epsilon).
\]
A short argument now shows that $\Pr[f \neq g] = O(\epsilon)$.

Since $g$ is obtained by majority decoding and each $g_S$ is a junta, we can show that $g$ is a sparse junta, proving \textbf{\Cref{itm:main-sparse}}. This allows us to show that $\EE_{\mu_p}[(f - g)^2] = O(\epsilon)$, a step which we highlight below.

Next, let us show why \textbf{\Cref{itm:main-non-boolean}} holds. Let $\bbone_S$ be a minimal non-Boolean input of $g$. Since $g$ is a sparse junta, the probability that $y_S = 1$ and $y_T = 0$ for all other minimal non-Boolean inputs $\bbone_T$ is $\Omega(p^{|S|})$. These events are disjoint, and so
\[
 \Pr[f \neq g] \geq \Pr[g \notin \{0,1\}] \geq \sum_S \Omega(p^{|S|}),
\]
where the sum goes over all minimal non-Boolean inputs of $g$. This implies \Cref{itm:main-non-boolean}.

Finally, let us address \textbf{\Cref{itm:main-boolean}}. The original function $g$ doesn't necessarily satisfy it. However, an argument along the lines of the preceding paragraph shows that the number of offending inputs is small, allowing us to slightly perturb $g$ so that it satisfies \Cref{itm:main-boolean}.

\paragraph{From $L_0$ to $L_2$ using the reverse union bound}

In the proof of \Cref{thm:main-intro}, we deduce the $L_2$ guarantee $\EE_{\mu_p}[(f - g)^2] = O(\epsilon)$ from the $L_0$ guarantee $\Pr[f \neq g] = O(\epsilon)$ using the fact that $g$ is a sparse junta. The argument uses a technical tool, the \emph{reverse union bound}, which we would like to highlight here.

Since $\Pr[f \neq g] = O(\epsilon)$ and $f$ is Boolean, it follows that $\Pr[g \notin \{0,1\}] = O(\epsilon)$, and so $\Pr[g(g-1) \neq 0] = O(\epsilon)$. We will deduce that $\EE_{\mu_p}[g^2(g-1)^2] = O(\epsilon)$, which implies $\EE_{\mu_p}[(f - g)^2] = O(\epsilon)$ via a short argument.

Since $g$ is a sparse junta whose monomial coefficients are quantized, the same holds for $h = g(g-1)$, and also for $h^2$. We will show that $\Pr[h \neq 0] = O(\epsilon)$ implies that $\EE_{\mu_p}[h^2] = O(\epsilon)$.

Since the monomial coefficients of $h^2$ are quantized, it suffices to show that the monomial support of $h^2$ is sparse, in the sense that it contains $O(\epsilon/p^e)$ sets of size $e$ for all $e \leq 2d$. Since $h$ is a sparse junta, in order to show that the monomial support of $h^2$ is sparse, it suffices to show that the monomial support of $h$ itself is sparse.

The union bound shows that
\[
 \Pr[h \neq 0] \leq \sum_{S \in \supp(h)} p^{|S|}.
\]
If the inequality were in the other direction then it would follow that the monomial support of $h$ is sparse, since $\Pr[h \neq 0] = O(\epsilon)$.
In general, we cannot reverse the inequality (consider for example $h = \sum_{i=1}^n y_i$, where the left-hand side tends to $1$ while the right-hand side tends to infinity for constant $p$).
However, when $h$ is a sparse junta, the \emph{reverse union bound} states that we can indeed reverse the direction of the inequality:
\[
 \Pr[h \neq 0] = \Omega\left(\sum_{S \in \supp(h)} p^{|S|}\right).
\]
The proof of the reverse union bound is not difficult. It uses the Harris--FKG inequality.

\subsection*{Differences from conference version}

This paper originates in an extended abstract~\cite{DinurFH2019}, which covers both the material eventually published in~\cite{DinurFH-agree} and part of the material in the present work. The full version of~\cite{DinurFH2019} is available on arXiv (1711.09428) and ECCC (TR17-180).

The full version of~\cite{DinurFH2019} contains weaker versions of \Cref{thm:main-intro,thm:main-monotone-intro}, which describe only some properties of the approximating function $g$, and in particular do not constitute a full characterization of all Boolean functions which are close to degree $d$. The current versions of these theorems are inspired by~\cite{Filmus2021}. In addition, \Cref{thm:junta-intro-formula} is completely new. It is inspired by~\cite{Filmus2023}.

In contrast, the main theorems in the full version of~\cite{DinurFH2019} apply in the more general \emph{$A$-valued} setting, which we describe briefly in \Cref{sec:junta}, as well as in the setting of the slice. We omit these generalizations in this version to make it more concise, and since these generalizations only require a few new ideas, the most interesting of which is that whereas the monomial expansion of functions on the slice is not unique, the sparse monomial expansion of a junta is unique.

\subsection*{Paper organization}

We start with various preliminaries in \Cref{sec:prel}. We formally define sparse juntas and describe some of their properties in \Cref{sec:sparsity}. We prove \Cref{thm:agreement-intro}, the junta agreement theorem, in \Cref{sec:agreement}. We prove the main structure theorem, \Cref{thm:main-intro}, in \Cref{sec:structure}, and its monotone version, \Cref{thm:main-monotone-intro}, in \Cref{sec:structure-monotone}. We prove the junta approximation theorem, \Cref{thm:junta-intro,thm:junta-intro-formula}, in \Cref{sec:junta}.

\section{Preliminaries}
\label{sec:prel}

We make constant use of the inequality $(a+b)^2 \leq 2a^2+2b^2$.

We sometimes use the notation $[n] = \{1,\dots,n\}$.

We use $\bbone_S$ for the characteristic vector of the set $S$ or for the indicator of the event $S$. We also use the notation $\indicator[S]$ for the latter to improve legibility.

The \emph{weight} of $y \in \{0,1\}^n$ is the number of coordinates equal to $1$.

A function on $\{0,1\}^n$ is an \emph{$M$-junta} if it depends on at most $M$ coordinates.

\paragraph{Monotone functions} If $x,y \in \{0,1\}^n$ then $x \leq y$ if $x_i \leq y_i$ for all $i \in [n]$. We write $x < y$ if $x \leq y$ and $x \neq y$.

A function $f\colon \{0,1\}^n \to \mathbb{R}$ is \emph{monotone} if $f(x) \leq f(y)$ whenever $x \leq y$.

If $f\colon \{0,1\}^n \to \{0,1\}$ is monotone then a \emph{minterm} of $f$ is an input $x$ such that $f(x) = 1$ and $f(y) = 0$ for all $y < x$. The set of minterms of $f$ constitutes its \emph{minterm support}.

A monotone function $f\colon \{0,1\}^n \to \{0,1\}$ is a \emph{width~$d$ monotone DNF} if all its minterms have weight at most~$d$.

\paragraph{Biased measure} For $n \in \mathbb{N}$ and $p \in \{0,1\}$, the measure $\mu_p$ on $\{0,1\}^n$ is given by
\[
 \mu_p(y_1,\dots,y_n) = p^{\sum_i y_i} (1-p)^{\sum_i (1-y_i)}.
\]
Equivalently, if we identify an element in $\{0,1\}^n$ with the corresponding subset, then a subset $S \sim \mu_p$ is sampled by including each element independently with probability $p$.

When the domain $\{0,1\}^n$ is not clear from context, we write it explicitly, for example $\mu_p(\{0,1\}^S)$ or $\mu_p(S)$.

Given a function $f\colon \{0,1\}^n \to \mathbb{R}$, we often think of the domain as endowed with the measure $\mu_p$ for some $p$. In this case, we write $f\colon (\{0,1\}^n, \mu_p) \to \mathbb{R}$. If we then write $\EE[f]$ or $\Pr[f \in \{0,1\}]$, then the underlying measure is $\mu_p$.

\paragraph{Closeness}

Our basic notion of closeness is $L_2$.
Two functions $f,g$ are \emph{$\epsilon$-close with respect to $\mu_p$} if $\EE_{\mu_p}[(f-g)^2] \leq \epsilon$. If $\mu_p$ is clear from context then we omit its mention.

\paragraph{Monomial expansion}
Every function $f\colon \{0,1\}^n \to \mathbb{R}$ has a unique representation as a multilinear polynomial:
\[
 f(y_1,\ldots,y_n) = \sum_{S \subseteq [n]} \tilde{f}(S) y_S, \text{ where } y_S = \prod_{i \in S} y_i,
\]
where $y_1,\ldots,y_n \in \{0,1\}$.
We call this representation the \emph{monomial expansion}. We stress that it differs from the Fourier expansion. 

The functions $y_S$ are called \emph{monomials}, and the coefficients $\tilde{f}(S)$ are called \emph{monomial coefficients}. The \emph{monomial support} of $f$ is
\[
 \supp(f) = \{ S \subseteq [n] : \tilde{f}(S) \neq 0 \}.
\]

The \emph{degree} of $f$, denoted $\deg f$, is the maximal size of a set in $\supp(f)$ (if $\supp(f) = \emptyset$ then $\deg f = 0$). This notion of degree coincides with the Fourier-theoretic notion of degree.

\paragraph{Low degree functions}

A function $f\colon \{0,1\}^n \to \mathbb{R}$ has \emph{degree $d$} if $\deg f \leq d$. A function is \emph{$\epsilon$-close to degree~$d$ (with respect to $\mu_p$)} if $\EE_{\mu_p}[(f - g)^2] \leq \epsilon$ for some degree~$d$ function $g$. 

We can also define these concepts via the $p$-biased Fourier expansion. Let $f = \sum_S \hat{f}(S) \omega_S$, where $\{ \omega_S \}$ is the $p$-biased Fourier basis, described in \Cref{sec:lem:expansion}. Let $f^{\leq d} = \sum_{|S| \leq d} \hat{f}(S) \omega_S$ and $f^{>d} = \sum_{|S| > d} \hat{f}(S) \omega_S$. The function $f$ has degree $d$ if $f^{>d} = 0$. The degree~$d$ function closest to $f$ (with respect to $\mu_p$) is $f^{\leq d}$, and so $f$ is $\epsilon$-close to degree~$d$ if $\EE_{\mu_p}[(f^{>d})^2] = \EE_{\mu_p}[(f - f^{\leq d})^2] \leq \epsilon$.

\paragraph{Function restriction}

We denote the restriction of $y \in \{0,1\}^n$ to the coordinates in a subset $S \subseteq [n]$ by $y|_S \in \{0,1\}^S$.

Given a function $f\colon \{0,1\}^n \to \mathbb{R}$ and a subset $S \subseteq [n]$, the restriction $f|_S\colon \{0,1\}^S \to \mathbb{R}$ is obtained by substituting zero for all coordinates outside of $S$:
\[
 f|_S(y) = \sum_{T \subseteq S} \tilde{f}(T) y_T.
\]

\paragraph{Function substitution}

Given a function $f\colon \{0,1\}^n \to \mathbb{R}$ and a subset $S \subseteq [n]$, the substitution $f|_{y_S \gets 1} \colon \{0,1\}^{\overline{S}} \to \mathbb{R}$ is obtained by substituting $y_i = 1$ for all $i \in S$:
\[
 f|_{y_S \gets 1} = \sum_T \tilde{f}(T) y_{T \setminus S}.
\]

If $S = \{i\}$ then we use the notation $f|_{y_i \gets 1}$.

\paragraph{Boolean functions}

A function is \emph{Boolean} if it is $\{0,1\}$-valued. If $f,g$ are Boolean then they are $\epsilon$-close if and only if $\Pr[f \neq g] \leq \epsilon$. In other words, for Boolean functions the $L_2$ and $L_0$ notions of distance coincide.

If $f\colon \{0,1\}^n \to \mathbb{R}$ then $\round(f, \{0,1\})$ is the Boolean function obtained by rounding each $f(y)$ to the nearest value among $\{0,1\}$. The original function $f$ is \emph{$\epsilon$-close to Boolean} if
\[
 \EE_y[\dist(f(y), \{0,1\})^2] \leq \epsilon,
\]
where $\dist(f(y), \{0,1\}) = \min(|f(y)-0|, |f(y)-1|)$. 

Equivalently, $\EE[(f - \round(f,\{0,1\}))^2] \leq \epsilon$.

\subsection{Unbiased structure theorems}
\label{sec:prel-structure}

\paragraph{Granularity}

The monomial coefficients of Boolean functions are bounded integers.

\begin{lemma} \label{lem:granularity}
If $f$ is Boolean and $S \neq \emptyset$ then $\tilde{f}(S)$ is an integer whose magnitude is at most $2^{|S|-1}$.
\end{lemma}
\begin{proof}
Let $g$ be the function obtained by substituting zero in all coordinates outside $S$. Then $\tilde{f}(S) = \tilde{g}(S)$ and
\[
 g(y) = \sum_{z \in \{0,1\}^S} f(z) \prod_{z_i = 0} (1 - y_i) \prod_{z_i = 1} y_i.
\]
Considering the coefficient of $y_S$, we see that
\[
 \tilde{f}(S) = \tilde{g}(S) = \sum_{z\colon z|_{\overline{S}} = 0} (-1)^{|S| - |z|} f(z).
\]
The claim follows since $2^{|S|-1}$ of the summands are $+f(z)$ and $2^{|S|-1}$ are $-f(z)$.
\end{proof}

When $S = \emptyset$, clearly $\tilde{f}(\emptyset) = f(0) \in \{0,1\}$.

\paragraph{Nisan--Szegedy}

Nisan and Szegedy~\cite{NisanS1994} proved the following theorem.

\begin{theorem}[Nisan--Szegedy] \label{thm:nisan-szegedy}
Every Boolean degree~$d$ function on $\{0,1\}^n$ is an $M_d$-junta, where $M_d = d2^{d-1}$.
\end{theorem}

The number of coordinates in the junta was improved to $O(2^d)$ in~\cite{ChiarelliHS2020,Wellens2022}.

\paragraph{Kindler--Safra}

In unpublished work, Kindler and Safra~\cite{KindlerS2002,Kindler2003} proved the following result, which we state in its formulation due to Keller and Klein~\cite{KellerK2020}.

\begin{theorem}[Kindler--Safra] \label{thm:kindler-safra-original}
There exists a universal constant $c > 0$ such that the following holds for all $d \in \mathbb{N}$.

If $f\colon (\{0,1\}^n, \mu_{1/2}) \to \{0,1\}$ is $\epsilon$-close to degree~$d$, where $\epsilon < c^d$, then $f$ is $2\epsilon$-close to a Boolean degree~$d$ function.
\end{theorem}

This formulation follows by combining Theorem~1.4 and Proposition~5.6 in~\cite{KellerK2020}. (Keller and Klein also prove a stronger version in which the closeness is improved to the optimal expression, $\epsilon + \tilde O(\epsilon^2)$.)

For our purposes, we need a version of the Kindler--Safra theorem which holds for all $\epsilon$.

\begin{theorem} \label{thm:kindler-safra}
There exists a universal constant $C > 0$ such that the following holds for all $d \in \mathbb{N}$.
If $f\colon (\{0,1\}^n, \mu_{1/2}) \to \{0,1\}$ is $\epsilon$-close to degree~$d$ then $f$ is $C^d\epsilon$-close to a Boolean degree~$d$ function.
\end{theorem}
\begin{proof}
We take $C = \max(1/c, 2)$, where $c$ is the constant in \Cref{thm:kindler-safra-original}.

Suppose that $f\colon (\{0,1\}^n, \mu_{1/2}) \to \{0,1\}$ is $\epsilon$-close to degree~$d$. If $\epsilon < c^d$ then the result follows directly from \Cref{thm:kindler-safra-original}. Otherwise, $C^d\epsilon \geq 1$, and so $f$ is $C^d\epsilon$-close to the constant zero function.
\end{proof}

\paragraph{Monotone Kindler--Safra}

When the function $f$ is monotone, we can guarantee that the approximating function in \Cref{thm:kindler-safra} is monotone as well.

We start with a warm-up.

\begin{lemma} \label{lem:non-monotone-junta}
Suppose that $f\colon (\{0,1\}^n, \mu_{1/2}) \to \mathbb{R}$ is a monotone function and $g\colon (\{0,1\}^n, \mu_{1/2}) \to \mathbb{R}$ is an $M$-junta.

Either $g$ is monotone or $\Pr[f \neq g] \geq 2^{-M}$.
\end{lemma}
\begin{proof}
Suppose without loss of generality that $g$ depends on the first $M$ coordinates. We separate accordingly the inputs to $f$ and $g$ into two parts: a point in $\{0,1\}^M$, and a point in $\{0,1\}^{n-M}$.

If $g$ is not monotone then $g(x,0) > g(y,0)$ for some $x < y$. For each $z \in \{0,1\}^{n-M}$ we have $g(x,z) > g(y,z)$ while $f(x,z) \leq f(y,z)$, and so either $f(x,z) \neq g(x,z)$ or $f(y,z) \neq g(y,z)$. Since there are $2^{n-M}$ choices for $z$,
\[
 \Pr[f \neq g] \geq \frac{2^{n-M}}{2^n} = 2^{-M}. \qedhere
\]
\end{proof}

Using a similar argument, we can prove a monotone version of the Kindler--Safra theorem. We are indebted to an anonymous reviewer for suggesting the following proof, which improves the bound on $K_d$ from doubly exponential to singly exponential.

\begin{theorem} \label{thm:kindler-safra-monotone}
For every $d \in \mathbb{N}$ there is a constant $K_d > 0$ such that the following holds.

If $f\colon (\{0,1\}^n, \mu_{1/2}) \to \{0,1\}$ is monotone and $\epsilon$-close to degree~$d$ then $f$ is $K_d\epsilon$-close to a monotone Boolean degree~$d$ function.
\end{theorem}
\begin{proof}
Suppose that $f\colon (\{0,1\}^n, \mu_{1/2}) \to \{0,1\}$ is monotone and $\epsilon$-close to degree~$d$. Applying \Cref{thm:kindler-safra}, $f$ is $C^d\epsilon$-close to a Boolean degree~$d$ function $g$.

If $g$ is monotone then we are done. Otherwise, there is an index $i$ and two inputs $x < y$ differing only in the $i$'th coordinate such that $g(x) > g(y)$. We will show that there are in fact many such pairs. To this end, define
\[
 g_i(y) = g(y|_{i \gets 1}) - g(y|_{i \gets 0}) \text{ and } G_i = \tfrac{1}{2} g_i (g_i - 1).
\]
Here $y|_{i\gets b}$ is obtained from $y$ by setting $y_i = b$.

The function $g_i$, which is just the derivative of $g$ with respect to $y_i$, attains the values $\{-1, 0, 1\}$, where $g_i(y) = -1$ if $(y_{i\gets 0},y_{i\gets 1})$ is a pair of inputs violating monotonicity. The function $G_i$ is the indicator function of this event.

By construction, $G_i$ has degree $2d-2$, and so $\EE[G_i]$ is an integer multiple of $2^{-2d-2}$. (One way to see this is to observe that the monomial coefficients of $G_i$ are integers, and that the expectation of a degree $e$ monomial is $2^{-e}$.) By construction, $\EE[G_i] > 0$, and so
\[
 \Pr[g(y_{i\gets 0}) = 1 \text{ and } g(y_{i\gets 1}) = 0] \geq \frac{1}{2^{2d-2}}.
\]
Since $f$ is monotone, this implies that
\[
 \Pr[f(y) \neq g(y)] \geq
 \frac{1}{2} \Pr[f(y_{i\gets 0}) \neq g(y_{i\gets 0}) \text{ or } f(y_{i\gets 1}) \neq g(y_{i\gets 1})] \geq \frac{1}{2^{2d-1}}.
\]
On the other hand, this probability is at most $C^d \epsilon$, and so $\epsilon \geq C^{-d}/2^{2d-1}$. It follows that $f$ is $2^{2d-1} C^d \epsilon$-close to the zero function.
\end{proof}

Curiously, the proof of the biased version of \Cref{thm:kindler-safra-monotone} only uses \Cref{lem:non-monotone-junta}.

\subsection{Forbidden configurations}
\label{sec:forbidden-configurations}

Our work involves three different types of \emph{forbidden configurations}. These are configurations of coefficients which cannot occur in Boolean functions, but do arise in our context --- but only sparingly. For example, the biased FKN theorem~\cite{Filmus2016-fkn} involves functions of the form $f = \sum_i c_i y_i$. If $f$ is Boolean then we cannot have $c_i = c_j = 1$, but if $f$ is $\epsilon$-close to Boolean then there could be up to $O(\epsilon/p^2)$ copies of this configuration.

\paragraph{Minimal non-Boolean inputs}
Let $f\colon \{0,1\}^n \to \mathbb{R}$. An input $x$ is a \emph{minimal non-Boolean input} of $f$ if $f(x) \notin \{0,1\}$ but $f(y) \in \{0,1\}$ for all $y < x$.

Suppose that $f$ has degree~$1$, say
\[
 f = c_0 + \sum_{i=1}^n c_i y_i.
\]

If $c_0 \notin \{0,1\}$ then $0$ is the only minimal non-Boolean input. Otherwise, suppose without loss of generality that $c_0 = 0$. If $c_i \notin \{0,1\}$ then $\bbone_{\{i\}}$ is a minimal non-Boolean input. If $c_i = c_j = 1$ then $\bbone_{\{i,j\}}$ is a minimal non-Boolean input. There are no other types of minimal non-Boolean inputs.

In addition, if $x$ is a minimal non-Boolean input then this is due to some non-zero monomial coefficients: if $\bbone_{\{i\}}$ is a minimal non-Boolean input then $c_i \neq 0$, and if $\bbone_{\{i,j\}}$ is a minimal non-Boolean input of $f$ then $c_i,c_j \neq 0$. In both cases, the minimal non-Boolean input $x = \bbone_S$ is such that $S$ is the union of sets in $\supp(f)$: $\{i\}$ in the former case, and $\{i\},\{j\}$ in the latter case.

The following lemma shows that a similar picture holds for all $d$. We thank an anonymous reviewer for suggesting the following proof, which improves the bound on $L_d$ from exponential to linear.

\begin{lemma} \label{lem:minimal-non-boolean}
For every $d$ there is a constant $L_d$ such that every minimal non-Boolean input of a degree~$d$ function $f\colon \{0,1\}^n \to \mathbb{R}$ has weight at most $L_d$.

Furthermore, if $\bbone_S$ is a minimal non-Boolean input of $f$ then $S$ can be written as the union of at most $L_d$ sets in $\supp(f)$.
\end{lemma}
\begin{proof}
Let $g = f (f - 1)$, and observe that $x$ is a minimal non-Boolean input of $f$ if and only if it is a minimal non-zero input of $g$, that is, $g(x) \neq 0$ but $g(y) = 0$ for all $y < x$. If $\bbone_S$ is a minimal non-zero input of $g$ then $\tilde{g}(S) \neq 0$, and so $|S| \leq \deg g \leq 2d$. Moreover, $S$ can be written as the union of at most two sets in $\supp(f)$.
\end{proof}

Looking ahead, we will be interested in minimal non-Boolean inputs of degree~$d$ functions whose weight is larger than $d$.
If $\bbone_S$ is a minimal non-Boolean input of $f$ then $f|_S$ is a function which is Boolean except for the input $\bbone_S$. Listing such \emph{minimal non-Boolean functions} is a convenient way to describe all minimal non-Boolean inputs. For example, when $d = 1$, the minimal non-Boolean functions are
\[
 y_1 + y_2, 1 - y_1 - y_2.
\]
In general, the minimal non-Boolean inputs come in pairs $f, 1-f$, and it suffices to list those which satisfy $f(0) = 0$.

Here are the minimal non-Boolean functions satisfying $f(0) = 0$ for $d = 2$: \begin{gather*}
    y_1 (y_2 + y_3) \\
    y_1 y_2 + y_1 y_3 + y_2 y_3 \\
    y_1 (1 - y_2 - y_3) \\
    y_1 + y_2 y_3 \\
    (y_1 + y_2)(1 - y_3) - 2y_1 y_2 \\
    y_1(1 - y_3) + y_2 - 2y_1 y_2 \\
    (y_1 + y_2)(1 - y_3) - y_1 y_2 \\
    y_1 + y_2 + y_3 - 2y_1 y_2 - 2y_1 y_3 - 2y_2 y_3 \\
    y_1 + y_2 + y_3 - 2y_1 y_2 - 2y_1 y_3 - y_2 y_3 \\
    y_1 + y_2 + y_3 - 2y_1 y_2 - y_1 y_3 - y_2 y_3 \\
    y_1 y_2 + y_3 y_4 \\
    y_1 (1 - y_3) + y_2 (1 - y_4) - y_1 y_2 \\
    y_1 + y_2 + y_3 - y_1 y_2 - y_1 y_3 - y_2 y_3 - y_1 y_4 \\
    y_1 + y_2 + y_3 + y_4 - y_1 y_2 - y_1 y_3 - y_1 y_4 - y_2 y_3 - y_2 y_4 - y_3 y_4
\end{gather*}

\paragraph{Minimal non-monotone inputs}

Let $f\colon \{0,1\}^n \to \{0,1\}$. An input $x$ is \emph{minimal non-monotone} if $f(x) < f(y)$ for some $y < x$, and $f(y) \leq f(z)$ whenever $y \leq z < x$.

\Cref{thm:nisan-szegedy} implies that each such input has weight at most $M_d$.

\begin{lemma} \label{lem:minimal-non-monotone}
Every minimal non-monotone input of a degree~$d$ function $f\colon \{0,1\}^n \to \{0,1\}$ has weight at most $M_d$, where $M_d$ is the constant from \Cref{thm:nisan-szegedy}.

Furthermore, if $\bbone_S$ is a minimal non-monotone input of $f$ then $S$ can be written as the union of at most $M_d$ sets in $\supp(f)$.
\end{lemma}
\begin{proof}
By minimality, if $\bbone_S$ is a minimal non-monotone input of $f$ then $f|_S$ depends on all inputs, and so $|S| \leq M_d$ by \Cref{thm:nisan-szegedy}. Moreover, $S$ is the union of all sets in $\supp(f|_S)$, and so it is the union of at most $|S|$ sets in $\supp(f)$.
\end{proof}

We extend the definition on minimal non-monotone inputs to non-Boolean functions as follows.
If $f\colon \{0,1\}^n \to \mathbb{R}$ then an input $x = \bbone_S$ is \emph{minimal non-monotone Boolean} if $f|_S$ is Boolean and $x$ is a minimal non-monotone input of $f|_S$.

\Cref{lem:minimal-non-monotone} holds also for minimal non-monotone Boolean inputs of a degree~$d$ function $g\colon \{0,1\}^n \to \mathbb{R}$, since if $\bbone_S$ is a minimal non-monotone Boolean input of $g$ then it is also a minimal non-monotone input of the Boolean function $g|_S$.

\smallskip

As in the case of minimal non-Boolean inputs, we will be interested in minimal non-monotone inputs of degree~$d$ functions whose weight is larger than $d$, which we can describe using \emph{minimal non-monotone functions}. When $d = 1$, there are no minimal non-monotone functions. When $d = 2$, the unique minimal non-monotone function is
\[
 y_1 + y_2 + y_3 - y_1 y_2 - y_1 y_3 - y_2 y_3.
\]

\paragraph{Minimal high-degree inputs}

Let $f\colon \{0,1\}^n \to \{0,1\}$, and let $d$ be a parameter. An input $\bbone_S$ is \emph{minimal high-degree} if $\deg(f|_S) > d$ but $\deg(f|_T) \leq d$ for all $T \subsetneq S$.

\begin{lemma} \label{lem:minimal-high-degree}
Every minimal high-degree input of a monotone width~$d$ DNF $f\colon \{0,1\}^n \to \{0,1\}$ has weight at most $L_d$, where $L_d$ is the constant from \Cref{lem:minimal-non-boolean}.

Furthermore, if $\bbone_S$ is a minimal high-degree input of $f$ then $S$ can be written as the union of at most $L_d$ minterms of $f$.
\end{lemma}
\begin{proof}
By minimality, if $\bbone_S$ is a minimal high-degree input of $f$ then $f|_S$ depends on all coordinates, and so $S$ is the union of at most $|S|$ minterms in $\supp(f|_S)$, say $B_1,\ldots,B_m$.

If $|S| \leq M_d$ then we are done. Otherwise, let $m$ be the minimal index such that $|B_1 \cup \cdots \cup B_m| > M_d$. By minimality, $\deg(f|_{B_1 \cup \cdots \cup B_{m-1}}) \leq d$, and so $|B_1 \cup \cdots \cup B_{m-1}| \le M_d$ by \Cref{thm:nisan-szegedy}. Since $f|_{B_1 \cup \cdots \cup B_m}$ depends on all inputs and $|B_1 \cup \cdots \cup B_m| > M_d$, \Cref{thm:nisan-szegedy} shows that it cannot have degree~$d$, and so $B_1 \cup \cdots \cup B_m = S$. Therefore $|S| \leq M_d + d$.
\end{proof}

As in the case of minimal non-Boolean inputs and minimal non-monotone inputs, we can describe minimal high-degree inputs as minimal high-degree monotone DNFs. When $d = 1$, the unique minimal high-degree monotone DNF is $y_1 \lor y_2$, and when $d = 2$, the minimal high-degree monotone DNFs are \begin{gather*}
    y_1 \land (y_2 \lor y_3) \\
    (y_1 \land y_2) \lor (y_1 \land y_3) \lor (y_2 \land y_3) \\
    y_1 \lor (y_2 \land y_3) \\
    y_1 \lor y_2 \lor y_3 \\
    (y_1 \land y_2) \lor (y_3 \land y_4)
\end{gather*}

\section{Sparsity}
\label{sec:sparsity}

The approximating function $g$ in \Cref{thm:main-intro} has sparse monomial support, and its monotone counterpart in \Cref{thm:main-monotone-intro} has sparse minterm support. This crucial property drives much of the proofs of these theorems.

Here is the notion of sparsity that comes up in both theorems.

\begin{definition}[Sparse]
\label{def:sparse}
A set system over a set $V$ is a collection of subsets of $V$.

A set system $\mathcal{F}$ is \emph{$(d,C,\epsilon)$-sparse}, where $d \ge 0$ is an integer, $\epsilon \ge 0$, and $C \ge 1$, if the following three properties hold:
\begin{enumerate}[(i)]
\item All sets in $\mathcal{F}$ have size at most $d$.
\item For all $T \subseteq [n]$ and integer $e \ge 0$, the set system $\mathcal{F}$ contains at most $C^e$ sets $U \supseteq T$ of size $|U| + e$.
\item For all integer $e \ge 0$, the set system $\mathcal{F}$ contains at most $C^e \epsilon$ sets $U$ of size $e$.
\end{enumerate}

A set system is $(d,C)$-sparse if it is $(d,C,1)$-sparse; this corresponds to omitting the final property.
\end{definition}

\begin{definition}[Sparse junta]
\label{def:sparse-junta}
A function $f\colon \{0,1\}^n \to \mathbb{R}$ is a \emph{$(d,C)$-sparse junta} if $\supp(f)$ is $(d,C)$-sparse.

A monotone function $f\colon \{0,1\}^n \to \{0,1\}$ is a \emph{$(d,C)$-sparse DNF} if its minterm support is $(d,C)$-sparse.
\end{definition}

We describe some elementary properties of sparse set systems in \Cref{sec:sparsity-elementary}. Sparse set systems satisfy a crucial property we call the \emph{reverse union bound}, which we explain in \Cref{sec:reverse-union-bound}. Finally, we discuss \emph{quantized} sparse juntas, which are sparse juntas in which the monomial coefficients belong to a fixed finite set. We show in \Cref{sec:quantized} that if a quantized sparse junta $f$ is $L_0$-close to Boolean (that is, $\Pr[f \notin \{0,1\}] \leq \epsilon$) then it is also $L_2$-close to Boolean (that is, $\EE[\dist(f, \{0,1\})^2] = O(\epsilon)$).

\subsection{Elementary properties}
\label{sec:sparsity-elementary}

We discuss the effect of three natural operations on the sparsity of set systems: union, sum and substitution.

The union of two set systems is their set-theoretic union, and it corresponds to the sum of functions:
\[
 \supp(f + g) \subseteq \supp(f) \cup \supp(g),
\]
where the inclusion is tight generically.

\begin{definition}[Sum of set systems]
\label{def:ssum}
If $\mathcal{F},\mathcal{G}$ are set systems then their \emph{sum} is
\[
 \mathcal{F} \ssum \mathcal{G} = \{ A \cup B : A \in \mathcal{F}, B \in \mathcal{G} \}.
\]

The \emph{$k'$th iterated sum} of a set system $\mathcal{F}$ is the sum of $k$ copies of $\mathcal{F}$:
\[
 \mathcal{F}^{\ssum k} = \{ A_1 \cup \cdots \cup A_k : A_1, \ldots, A_k \in \mathcal{F} \}.
\]
Since $A_1,\ldots,A_k$ are not necessarily distinct, $\mathcal{F}^{\ssum k}$ includes $\mathcal{F}^{\ssum \ell}$ for all non-zero $\ell < k$.
\end{definition}

The sum of two set systems corresponds to the product of functions:
\[
 \supp(fg) \subseteq \supp(f) \ssum \supp(g),
\]
where the inclusion is tight generically.

\begin{definition}[Contraction]
If $\mathcal{F}$ is a set system over $V$ and $S \subseteq V$ then
\[
 \mathcal{F}|_{S \gets 1} = \{ T \setminus S : T \in \mathcal{F} \}.
\]
\end{definition}

Contraction corresponds to substituting the value of variables:
\[
 \supp(f|_{y_S \gets 1}) \subseteq \supp(f)|_{S \gets 1},
\]
and the inclusion is tight generically.

Union, sum and contraction all preserve sparsity, after adapting the parameters.

\begin{lemma} \label{lem:sparsity-elementary}
Let $\mathcal{F},\mathcal{G}$ be $(d,C)$-sparse set systems.
\begin{enumerate}[(a)]
\item $\mathcal{F} \cup \mathcal{G}$ is $(d,2C)$-sparse.
\label{itm:sparsity-o}
\item $\mathcal{F} \ssum \mathcal{G}$ is $(2d,O_d(C))$-sparse.
\label{itm:sparsity-a}
\item $\mathcal{F}^{\ssum k}$ are $(kd,O_{k,d}(C))$-sparse.
\label{itm:sparsity-b}
\item $\mathcal{F}|_{J \gets 1}$ is $(d,O_{d,|J|}(C))$-sparse.
\label{itm:sparsity-c}
\end{enumerate}
\end{lemma}
\begin{proof}
Every set system satisfies the definition of $(d,C)$-sparse for $e = 0$, and so it suffices to prove it for $e \geq 1$.

\paragraph{\Cref{itm:sparsity-o}.} Let $\mathcal{F},\mathcal{G}$ be $(d,C)$-sparse set systems. Given $R$ and $e \geq 1$, there are at most $C^e$ sets in each of $\mathcal{F},\mathcal{G}$ of size $|R|+e$ which contain $R$. Therefore there are at most $2C^e \le (2C)^e$ sets in $\mathcal{F} \cup \mathcal{G}$ of size $|R|+e$ which contain $R$.

\paragraph{\Cref{itm:sparsity-a}.} Let $\mathcal{F},\mathcal{G}$ be $(d,C)$-sparse set systems. Given $R$ and $e$, we need to bound the number of sets $S = A \cup B$ such that $A \in \mathcal{F}$, $B \in \mathcal{G}$, $A \cup B \supseteq R$, and $|A \cup B| = |R| + e$. We can assume that $|R| \leq 2d$.

We will show that for every $a,a_r,a_b \leq |R|$ there are at most $16^dC^e$ pairs $(A,B)$ such that:
\begin{enumerate}[(i)]
\item $A \in \mathcal{F}$, $B \in \mathcal{G}$, $A \cup B \supseteq R$, $|A \cup B| = |R| + e$.
\item $|A| = a$, $|A \cap R| = a_r$, and $|A \cap B| = a_b$.
\end{enumerate}
Since there are at most $(2d+1)^3$ many options for $a,a_r,a_b$, this will complete the proof.

There are $\binom{|R|}{a_r} \leq 4^d$ choices for $A \cap R$ such that $|A \cap R| = a_r$. Since $\mathcal{F}$ is $(d,C)$-sparse, there are at most $C^{a - a_r}$ sets $A \in \mathcal{F}$ extending $A \cap R$ of size $a$. There are $\binom{|A|}{a_b} \leq 2^a \leq 4^d$ choices for $A \cap B$.

Notice that $B$ needs to contain both $A \cap B$ and $R \setminus A$, two sets which are disjoint. Also, since $|A \cup B| = |R| + e$, we have $|B| = |A \cup B| + |A \cap B| - |A| = |R| + e + a_b - a$.
Since $\mathcal{G}$ is $(d,C)$-sparse, the number of sets $B \in \mathcal{G}$ of size $b$ extending $(A \cap B) \cup (R \setminus A)$ is at most $C^{(|R| + e + a_b - a) - a_b - (|R|-a_r)} = C^{e - a + a_r}$.

Altogether, given $a,a_r,a_b$, the number of pairs $A,B$ satisfying the properties listed above is bounded by
\[
\underbrace{4^d}_{A \cap R} \cdot \underbrace{C^{a - a_r}}_{A \text{ given } A \cap R} \cdot \underbrace{4^d}_{A \cap B \text{ given } A} \cdot \underbrace{C^{e - a + a_r}}_{B \text{ given } A \cap R, A \cap B} = 16^d C^e,
\]
as required.

\paragraph{\Cref{itm:sparsity-b}.} This item follows from \Cref{itm:sparsity-a} by induction on $k$.

\paragraph{\Cref{itm:sparsity-c}.} Let $\mathcal{F}$ be a $(d,C)$-sparse set system, and let $J$ be a set. Given $R$ and $e$, we need to bound the number of sets $T \in \mathcal{F}$ such that $T \setminus J \supseteq R$ and $|T \setminus J| = |R| + e$. We can assume that $R$ is disjoint from $J$.

We will bound the number of such sets $T$ given $I = T \cap J$. Any such set $T$ contains both $I$ and $R$, and $|T| = |T \setminus J| + |T \cap J| = |R| + e + |I|$. Since $\mathcal{F}$ is $(d,C)$-sparse, there are at most $C^{(|R| + e + |I|) - (|I| + |R|)} = C^e$ sets $T \in \mathcal{F}$ containing $I \cup R$ and of size $|R| + e + |I|$. We conclude that the number of sets $T \in \mathcal{F}$ such that $T \setminus J \supseteq R$ and $|T \setminus J| = |R| + e$ is at most
\[
 \sum_{I \subseteq J} C^e \leq 2^{|J|} C^e \leq (2^{|J|}C)^e. \qedhere
\]
\end{proof}

\begin{lemma} \label{lem:sparsity-elementary-eps}
Let $\mathcal{F},\mathcal{G}$ be $(d,C,\epsilon)$-sparse.

\begin{enumerate}[(a)]
\item $\mathcal{F} \cup \mathcal{G}$ is $(d,2C,\epsilon)$-sparse.
\label{itm:sparsity-eps-o}
\item $\mathcal{F} \ssum \mathcal{G}$ is $(2d,O_d(C),\epsilon)$-sparse. For this it suffices that $\mathcal{G}$ be $(d,C)$-sparse.
\label{itm:sparsity-eps-a}
\item $\mathcal{F}^{\ssum k}$ are $(kd,O_{k,d}(C),\epsilon)$-sparse.
\label{itm:sparsity-eps-b}
\end{enumerate}
\end{lemma}
\begin{proof}
~\paragraph{\Cref{itm:sparsity-eps-o}.} Let $\mathcal{F},\mathcal{G}$ be $(d,C,\epsilon)$-sparse set systems. \Cref{lem:sparsity-elementary} shows that $\mathcal{F} \cup \mathcal{G}$ is $(d,2C)$-sparse. Given $e \geq 1$, there are at most $C^e \epsilon$ sets of size $e$ in each of $\mathcal{F}, \mathcal{G}$. Therefore there are at most $2C^e \epsilon \leq (2C)^e \epsilon$ sets of size $e$ in $\mathcal{F} \cup \mathcal{G}$.

The case $e = 0$ requires special treatment. If $\epsilon < 1$ then $\mathcal{F},\mathcal{G}$ do not contain $\emptyset$, and so $\mathcal{F} \cup \mathcal{G}$ also doesn't contain $\emptyset$. If $\epsilon \geq 1$ then clearly $\mathcal{F} \cup \mathcal{G}$ contains at most $\epsilon$ sets of size~$0$.

\paragraph{\Cref{itm:sparsity-eps-a}.} Let $\mathcal{F}$ be a $(d,C,\epsilon)$-sparse set system and let $\mathcal{G}$ be a $(d,C)$-sparse set system. \Cref{lem:sparsity-elementary} shows that $\mathcal{F} \ssum \mathcal{G}$ is $(d,O_d(C))$-sparse.

Given $e$, we need to bound the number of sets $S = A \cup B$ of size $e$, where $A \in \mathcal{F}$ and $B \in \mathcal{G}$. For this, we show that for each $a,a_b \leq d$, there are at most $2^d C^e \epsilon$ sets $A \cup B$ of size $e$ such that $A \in \mathcal{F}$, $B \in \mathcal{G}$, $|A| = a$, and $|A \cap B| = a_b$.

Indeed, there are at most $C^a \epsilon$ sets $A \in \mathcal{F}$ of size $a$. Given $A$, there are at most $\binom{|A|}{a_b} \leq 2^d$ choices for $A \cap B$. Since $|A \cup B| = e$, we have $|B| = |A \cup B| + |A \cap B| - |A| = e + a_b - a$. Therefore given $A \cap B$, there are at most $C^{e - a}$ sets $B \in \mathcal{G}$ including $A \cap B$ such that $|A \cup B| = e$.
Altogether, given $a, a_b$, the number of pairs $A, B$ satisfying the properties listed above is bounded by
\[
 \underbrace{C^a \epsilon}_A \cdot \underbrace{2^d}_{A \cap B \text{ given } A} \cdot \underbrace{C^{e - a}}_{B \text{ given } A \cap B} = 2^d C^e \epsilon,
\]
as claimed.

Since there are at most $(d+1)^2$ many options for $a, a_b$, this shows that the number of sets $S \in \mathcal{F} \ssum \mathcal{G}$ of size $e \geq 1$ is at most $(d+1)^2 2^d C^e \epsilon \leq ((d+1)^2 2^d C)^e \epsilon$, completing the proof except for the case $e = 0$.

In order to handle the case $e = 0$, we again consider whether $\epsilon < 1$ or not. If $\epsilon < 1$ then $\emptyset \notin \mathcal{F}$ and so $\emptyset \notin \mathcal{F} \ssum \mathcal{G}$. If $\epsilon \geq 1$ then $\mathcal{F} \ssum \mathcal{G}$ trivially contains at most $\epsilon$ sets of size~$0$.

\paragraph{\Cref{itm:sparsity-eps-b}.} This item follows from \Cref{itm:sparsity-eps-a} by induction on $k$.
\end{proof}

\subsection{Reverse union bound}
\label{sec:reverse-union-bound}

Let $\mathcal{F}$ be a set system. The union bound shows that
\[
 \Pr_{Y \sim \mu_p}[Y \supseteq S \text{ for some } S \in \mathcal{F}] \leq \sum_{S \in \mathcal{F}} p^{|S|}.
\]
In general, the union bound is not tight. For example, suppose that $\mathcal{F} = \{ \{1\}, \ldots, \{n\} \}$. As $n \to \infty$, the left-hand side tends to $1$ while the right-hand side tends to infinity. More generally, if $\mathcal{F} = \{ [m] \cup \{m+1\}, \ldots, [m] \cup \{n\} \}$ then the left-hand side tends to $p^m$ while the right-hand side tends to infinity.

In this section, we show that the union bound is tight up to a constant factor provided that $\mathcal{F}$ is sparse.

\begin{lemma}[Reverse union bound] \label{lem:reverse-union-bound}
Let $p \leq 1/2$, let $\mathcal{G}$ be a $(d,C/p)$-sparse set system, and let $\mathcal{F} \subseteq \mathcal{G}$.

For $S \in \mathcal{F}$, let $\mathcal{E}_S$ denote the event that $y_S = 1$ and $y_R = 0$ for all $R \in \mathcal{G}$ such that $R \not\subseteq S$.

Let $\mathcal{E}$ denote the union of the events $\mathcal{E}_S$ for all $S \in \mathcal{F}$. Then
\[
 \Pr[\mathcal{E}] =
 \Theta_{d,C}\left(
 \sum_{S \in \mathcal{F}} p^{|S|}
 \right).
\]
\end{lemma}
\begin{proof}
If $\mathcal{E}$ holds then $y_S = 1$ for some $S \in \mathcal{F}$, and so
\[
 \Pr[\mathcal{E}] \leq \sum_{S \in \mathcal{F}} p^{|S|}.
\]

In the other direction, observe first that the events $\mathcal{E}_S$ are disjoint. Indeed, suppose that $S,T \in \mathcal{F}$ are distinct. If $S \subseteq T$ and $T \subseteq S$ then $S = T$, so assume without loss of generality that $S \not\subseteq T$. If $\mathcal{E}_S$ holds then $y_S = 1$ and so $\mathcal{E}_T$ cannot hold. This implies that
\[
 \Pr[\mathcal{E}] = \sum_{S \in \mathcal{F}} \Pr[\mathcal{E}_S].
\]
We now bound each of the summands:
\[
 \Pr[\mathcal{E}_S] = p^{|S|} \Pr[y_T = 0 \text{ for all } \emptyset \neq T \in \mathcal{G}|_{S \gets 1}] \stackrel{(*)}\ge p^{|S|} \prod_{\emptyset \neq T \in \mathcal{G}|_{S \gets 1}} \Pr[y_T = 0],
\]
where $(*)$ follows from the Harris--FKG inequality, since the events $y_T = 0$ are anti-monotone. 
\Cref{lem:sparsity-elementary} shows that $\mathcal{G}|_{S \gets 1}$ is $(d,K/p)$-sparse, where $K$ depends only on $d,C$. Therefore
\[
 \Pr[\mathcal{E}_S] \geq p^{|S|} \prod_{e=1}^d (1-p^e)^{K^e/p^e} \geq 
 p^{|S|} \prod_{e=1}^d (1-2^{-e})^{(2K)^e} = \Omega_{d,C}(p^{|S|}),
\]
since $(1-x)^{1/x}$ is decreasing in $[0,1]$.

We conclude that
\[
 \Pr[\mathcal{E}] = \sum_{S \in \mathcal{F}} \Pr[\mathcal{E}_S] = \Omega_{d,C}\left( \sum_{S \in \mathcal{F}} p^{|S|} \right). \qedhere
\]
\end{proof}

When $\mathcal{F} = \mathcal{G}$, we recover the statement that the union bound is tight up to a constant factor for sparse set systems. Indeed, if $\mathcal{E}$ happens then in particular $Y \supseteq S$ for some $S \in \mathcal{F}$, and so the \namecref{lem:reverse-union-bound} shows that
\[
 \Pr_{Y \sim \mu_p}[Y \supseteq S \text{ for some } S \in \mathcal{F}] \geq \Pr[\mathcal{E}] = \Omega_{d,C}\left(\sum_{S \in \mathcal{F}} p^{|S|}\right).
\]

\subsection{Quantized sparse juntas}
\label{sec:quantized}

The sparse juntas that we consider later on will arise from applying the junta agreement theorem, \Cref{thm:agreement-intro}. Such functions are \emph{quantized} in the following sense.

\begin{definition}[Quantized functions] \label{def:quantized}
A function $f\colon \{0,1\}^n \to \mathbb{R}$ is \emph{$B$-quantized}, where $B$ is a finite set, if all monomial coefficients of $f$ belong to $B$.
\end{definition}

Quantization is preserved under basic operations.

\begin{lemma} \label{lem:quantized}
Let $f,g\colon \{0,1\}^n \to \mathbb{R}$ be $B$-quantized functions of degree~$d$, where $B$ is a finite set.

\begin{enumerate}[(a)]
\item For any $c \in \mathbb{R}$, the function $cf$ is $cB$-quantized, where $cB = \{ cb : b \in B \}$.
\label{itm:quantized-o}
\item The function $f + g$ is $(B + B)$-quantized, where $B + B = \{ b_1 + b_2 : b_1, b_2 \in B \}$.
\label{itm:quantized-a}
\item The function $fg$ is $C$-quantized, where $C$ is a finite set depending only on $d,B$.
\label{itm:quantized-b}
\end{enumerate}
\end{lemma}
\begin{proof}
\Cref{itm:quantized-o,itm:quantized-a} are clear. For \Cref{itm:quantized-b}, we use the formula
\[
 \widetilde{fg}(U) = \sum_{S \cup T = U} \tilde{f}(S) \tilde{g}(T),
\]
which shows that $\widetilde{fg}(U)$ is the sum of $3^{|U|} \leq 3^{2d}$ products of two elements from $B$.
\end{proof}

The junta agreement theorem will result in a quantized sparse junta $g$ which is close to Boolean in an $L_0$ sense: $\Pr[g \notin \{0,1\}] = O(\epsilon)$. In this section, we  show how to conclude closeness in an $L_2$ sense: $\Pr[\dist(g, \{0,1\})^2] = O(\epsilon)$.

We start with a similar result which compares the $L_0$ and $L_2$ norms.

\begin{lemma} \label{lem:L0-L2-norm}
If $f\colon (\{0,1\}^n, \mu_p) \to \mathbb{R}$ is a $B$-quantized $(d,C/p)$-sparse junta, where $p \leq 1/2$, then
\[
 \Pr[f \neq 0] = \Theta_{d,C,B}(\EE[f^2]).
\]
\end{lemma}
\begin{proof}
If $f \neq 0$ then clearly $y_S = 1$ for some $S \in \supp(f)$, and so the union bound shows that
\[
 \Pr[f \neq 0] \leq \sum_{S \in \supp(f)} p^{|S|}.
\]
We can get a matching lower bound using \Cref{lem:reverse-union-bound}. Let $\mathcal{F}$ be all inclusion-minimal sets in $\supp(f)$, and let $\mathcal{G} = \supp(f)$. If the event $\mathcal{E}_S$ in the \namecref{lem:reverse-union-bound} happens for some $S \in \mathcal{F}$ then $f = \tilde{f}(S) \neq 0$, and so
\[
 \Pr[f \neq 0] \geq \Pr[\mathcal{E}] = \Omega_{d,C} \left(\sum_{S \in \mathcal{F}} p^{|S|}\right).
\]
Since $f$ is a $(d,C/p)$-sparse junta,
\[
 \sum_{S \in \supp(f)} p^{|S|} = 
 \sum_{S \in \mathcal{F}} p^{|S|} \sum_{e=0}^{d-|S|} \sum_{\substack{T \in \supp(f) \\ T \supseteq S, |T| = |S| + e}} p^e \leq
 \sum_{S \in \mathcal{F}} p^{|S|} \sum_{e=0}^{d-|S|} (C/p)^e p^e = O_{d,C}\left(\sum_{S \in \mathcal{F}} p^{|S|}\right),
\]
and we conclude that
\begin{equation} \label{eq:L0-L2-norm}
 \Pr[f \neq 0] \geq \Pr[\mathcal{E}] = \Omega_{d,C} \left(\sum_{S \in \supp(f)} p^{|S|}\right).
\end{equation}
Similarly,
\[
 \EE[f^2] \geq \min_{\substack{b \in B \\ b \neq 0}} b^2 \Pr[\mathcal{E}] = \Omega_{d,C,B} \left(\sum_{S \in \supp(f)} p^{|S|}\right) = \Omega_{d,C,B}(\Pr[f \neq 0]), 
\]
showing that $\Pr[f \neq 0] = O_{d,C,B}(\EE[f^2])$.

For the other direction, \Cref{eq:L0-L2-norm} implies that $\supp(f)$ is $(d,C/p,\epsilon)$-sparse, where $\epsilon = O_{d,C}(\Pr[f \neq 0])$.
\Cref{lem:sparsity-elementary-eps} shows that $\supp(f^2) \subseteq \supp(f)^{\ssum 2}$ is $(2d,K/p,\epsilon)$-sparse, where $K = O_{d,C}(1)$, and \Cref{lem:quantized} shows that $f^2$ is $B'$-quantized for some finite $B'$ depending only on $d,B$. Therefore
\[
 \EE[f^2] \leq \max_{b \in B'} b \sum_{e = 0}^{2d} (K/p)^e \epsilon \cdot p^e = O_{d,C,B}(\Pr[f \neq 0]). \qedhere
\]
\end{proof}

Using this, we derive the main result of this section.

\begin{lemma} \label{lem:L0-L2}
If $f\colon (\{0,1\}^n, \mu_p) \to \mathbb{R}$ is a $B$-quantized $(d,C/p)$-sparse junta, where $p \leq 1/2$, then
\[
 \Pr[f \notin \{0,1\}] = \Theta_{d,C,B}(\EE[\dist(f, \{0,1\})^2]).
\]
\end{lemma}
\begin{proof}
Let $g = f(f-1)$. Since $\supp(g) \subseteq \supp(f)^{\ssum 2}$, \Cref{lem:sparsity-elementary,lem:quantized} show that $g$ is a $B'$-quantized $(2d,K/p)$-sparse junta, where $B',K$ depend on $d,C,B$. Therefore, \Cref{lem:L0-L2-norm} implies that
\[
 \Pr[f \notin \{0,1\}] = \Pr[g \neq 0] = \Theta_{d,C,B}(\EE[g^2]).
\]

We show that $\EE[\dist(f, \{0,1\})^2] = O_{d,C,B}(\Pr[f \notin \{0,1\}])$ by relating $\EE[\dist(f, \{0,1\})^2]$ to $\EE[g^2]$.
Consider any input $y$. If $\round(f(y), \{0,1\}) = a$ then $|f(y) - (1-a)| \geq 1/2$ and so $(f(y)-(1-a))^2 \geq 1/4$. Consequently,
\[
 \dist(f(y), \{0,1\})^2 = (f(y)-a)^2 \leq 4(f(y)-a)^2(f(y)-(1-a))^2 = 4g(y)^2.
\]
It follows that
\[
 \EE[\dist(f, \{0,1\})^2] \leq 4\EE[g^2] = O_{d,C,B}(\Pr[f \notin \{0,1\}]).
\]

\smallskip

The proof of the other direction, $\Pr[f \notin \{0,1\}] = O_{d,C,B}(\EE[\dist(f, \{0,1\})^2])$, uses the notion of minimal non-Boolean input, defined in \Cref{sec:forbidden-configurations}.
If $f(y) \notin \{0,1\}$ then $y \geq \bbone_S$ for some minimal non-Boolean input $\bbone_S$ of $f$, implying that $y_S = 1$. Denoting the set of minimal non-Boolean inputs of $f$ by $\mathcal{F}$, this shows that
\[
 \Pr[f \notin \{0,1\}] \leq \sum_{\bbone_S \in \mathcal{F}} \Pr[y_S = 1] = \sum_{\bbone_S \in \mathcal{F}} p^{|S|}.
\]

We would like to bound the right-hand side using \Cref{lem:reverse-union-bound}, but it is not necessarily the case that $\mathcal{F} \subseteq \supp(f)$. However, \Cref{lem:minimal-non-boolean} shows that $\mathcal{F} \subseteq \supp(f)^{\ssum L_d}$. \Cref{lem:sparsity-elementary} shows that $\supp(f)^{\ssum L_d}$ is $(L_dd,K'/p)$-sparse, where $K'$ depends on $d,C$. Therefore, \Cref{lem:reverse-union-bound}, applied with $\mathcal{F}$ and $\mathcal{G} = \supp(f)^{\ssum L_d}$, shows that
\[
 \Pr[f \notin \{0,1\}] \leq \sum_{S \in \mathcal{F}} p^{|S|} = \Theta_{d,C}(\Pr[\mathcal{E}]),
\]
where $\mathcal{E}$ is the event that $y_S = 1$ for some $S \in \mathcal{F}$, and $y_R = 0$ for all $R \in \supp(f)^{\ssum L_d}$ such that $R \not\subseteq S$. If the event $\mathcal{E}$ happens then $f(y) = f(\bbone_S) \notin \{0,1\}$. Moreover, since $|S| \leq L_d$ by \Cref{lem:minimal-non-boolean} and $f$ is $B$-quantized, $\dist(f(y), \{0,1\})^2 = \Omega_{d,B}(1)$. Therefore,
\[
 \Pr[f \notin \{0,1\}] = O_{d,C,B}(\EE[\dist(f, \{0,1\})^2]). \qedhere
\]
\end{proof}

\section{Junta agreement theorem}
\label{sec:agreement}

We prove our main structure theorems, \Cref{thm:main-intro,thm:main-monotone-intro}, by reduction to the Kindler--Safra theorem on $(\{0,1\}^n, \mu_{1/2})$. As we explain in \Cref{sec:intro-proof-sketch}, this involves applying the Kindler--Safra theorem to $f|_S$ for $S \sim \mu_{2p}$, obtaining approximating juntas $g_S$, and pasting the juntas together to a global function $g$.

In order to show that the juntas $g_S$ can be pasted together, we need to assume that they agree with each other, in the sense that
\[
 \Pr_{(S_1,S_2,T) \sim \nu} [g_{S_1}|_T \neq g_{S_2}|_T] \leq \epsilon
\]
for an appropriate distribution $\nu$ supported on triplets $(S_1,S_2,T)$ such that $S_1,S_2 \supseteq T$.

Since we are interested in the behavior of $g_S$ for $S \sim \mu_{2p}$, we need the marginals of $S_1$ and~$S_2$ to be $\mu_{2p}$. Moreover, we want $S_1$ and $S_2$ to be independent given $T$. This naturally leads to the product distribution $\mu_{q,r}$ (where $q = 2p$ and $r \leq q$), which is defined as follows:
\begin{enumerate}
    \item $T$ includes each $i \in [n]$ with probability $r$.
    \item For $\ell \in \{1,2\}$, $S_\ell$ includes each $i \in T$ with probability $1$, and each $i \notin T$ with probability $\frac{q-r}{1-r}$.
\end{enumerate}
It is easy to check that the marginal distributions of $S_1,S_2,T$ are $\mu_q,\mu_q,\mu_r$, respectively.

We can now state the junta agreement theorem, which is a formal statement of \Cref{thm:agreement-intro}.

\begin{theorem}[Junta agreement theorem] \label{thm:agreement}
Fix an integer $K$ and $c \in (0,1)$. The following holds for all $q \in (0,1)$.

Suppose that for each $S \subseteq [n]$ we are given a degree~$d$ function $g_S\colon \{0,1\}^n \to \{0,1\}$ whose monomial support contains at most $K$ monomials (the \emph{junta assumption}). Assume that the functions $g_S$ satisfy the following \emph{agreement condition}:
\[
 \Pr_{(S_1,S_2,T) \sim \mu_{q,cq}}[g_{S_1}|_T \neq g_{S_2}|_T] \leq \epsilon.
\]

Define a degree~$d$ function $g\colon \{0,1\}^n \to \{0,1\}$ as follows: for each $|A| \leq d$, let $\tilde{g}(A)$ be any value $a$ maximizing
\[
 \Pr_{\substack{S \sim \mu_q \\ S \supseteq A}}[\tilde{g}_S(A) = a].
\]

The function $g$ is a $(d,O_{d,K,c}(1/q))$-sparse junta satisfying
\[
 \Pr_{S \sim \mu_q}[g|_S \neq g_S] = O_{d,K,c}(\epsilon).
\]
\end{theorem}

In the statement of the \namecref{thm:agreement} and below, we use the convention that in $\Pr$ and $\EE$, we take the distribution on the first line of the subscript conditioned on the constraints in the following lines. For example, in the second display, $S$ is chosen by including all elements in $A$ with probability $1$ and all elements outside of $A$ with probability $q$.

\smallskip

Our previous work~\cite{DinurFH-agree} proves a stronger version of \Cref{thm:agreement}, which does not require the $g_S$ to be juntas but has the same conclusion (without the guarantee that $g$ is a sparse junta, which follows from the $g_S$ being juntas). The proof of this stronger version is somewhat non-intuitive, using the notion of good and excellent sets which is borrowed from~\cite{ImpagliazzoKW2012}. It turns out that the proof simplifies dramatically when the $g_S$ are juntas, and this is the proof that we present here.

The proof of \Cref{thm:agreement} relies on the following expansion lemma.

\begin{lemma} \label{lem:expansion}
Assume the setup of \Cref{thm:agreement}. There exists a constant $\expT_c$, depending only on $c$, such that for every $|A| \leq d$,
\[
 \Pr_{\substack{S_1, S_2 \sim \mu_q \\ S_1, S_2 \supseteq A}}[\tilde{g}_{S_1}(A) \neq \tilde{g}_{S_2}(A)] \leq \expT_c
 \Pr_{\substack{(S_1,S_2,T) \sim \mu_{q,cq} \\ T \supseteq A}}[\tilde{g}_{S_1}(A) \neq \tilde{g}_{S_2}(A)].
\]
\end{lemma}

We first derive \Cref{thm:agreement} assuming \Cref{lem:expansion}, and then prove \Cref{lem:expansion} using elementary Fourier analysis.

\begin{proof}[Proof of Theorem \ref{thm:agreement}]
~\paragraph{Agreement}
Our first goal is to show that $\Pr[g|_S \neq g_S] = O_{K,c}(\epsilon)$. We start by applying the union bound to relate the probability to disagreement on a particular monomial coefficient:
\[
 \Pr_{S \sim \mu_q}[g|_S \neq g_S] \leq
 \sum_{|A| \leq d} \Pr_{S \sim \mu_q}[S \supseteq A \text{ and } \tilde{g}(A) \neq \tilde{g}_S(A)] =
 \sum_{|A| \leq d} q^{|A|} \Pr_{\substack{S \sim \mu_q \\ S \supseteq A}}[\tilde{g}(A) \neq \tilde{g}_S(A)].
\]
We chose $\tilde{g}(A)$ as a value $a$ minimizing $\Pr[\tilde{g}_S(A) \neq a \mid S \supseteq A]$, and so
\[
 \Pr_{S \sim \mu_q}[g|_S \neq g_S] \leq
 \sum_{|A| \leq d} q^{|A|} \Pr_{\substack{S_1,S_2 \sim \mu_q \\ S \supseteq A}}[\tilde{g}_{S_1}(A) \neq \tilde{g}_{S_2}(A)].
\]
We can relate this to the agreement condition using \Cref{lem:expansion}:
\begin{align*}
 \Pr_{S \sim \mu_q}[g|_S \neq g_S] &\leq \expT_c
 \sum_{|A| \leq d} q^{|A|}
 \Pr_{\substack{(S_1,S_2,T) \sim \mu_{q,cq} \\ T \supseteq A}}[\tilde{g}_{S_1}(A) \neq \tilde{g}_{S_2}(A)] \\ &= \expT_c
 \sum_{|A| \leq d} c^{-|A|}
 \Pr_{(S_1,S_2,T) \sim \mu_{q,cq}}[T \supseteq A \text{ and } \tilde{g}_{S_1}(A) \neq \tilde{g}_{S_2}(A)].
\end{align*}
In order to use the junta assumption, we move the sum over $A$ inside:
\[
 \Pr_{S \sim \mu_q}[g|_S \neq g_S] \leq \expT_c c^{-d}
 \EE_{(S_1,S_2,T) \sim \mu_{q,cq}}[\operatorname{disagreements}(g_{S_1}|_T, g_{S_2}|_T)],
\]
where $\operatorname{disagreements}$ counts the number of different monomial coefficients. The junta assumption implies that there are at most $2K$ disagreements, and so
\[
 \Pr_{S \sim \mu_q}[g|_S \neq g_S] \leq \expT_c c^{-d} 2K \Pr_{(S_1,S_2,T) \sim \mu_{q,cq}}[g_{S_1}|_T \neq g_{S_2}|_T] = O_{d,K,c}(\epsilon).
\]

\paragraph{Sparse junta}
It remains to show that $g$ is a $(d,O_{K,c}(1/q))$-sparse junta. The main observation is that if $\tilde{g}(A) \neq 0$ then
\[
 \Pr_{\substack{S \sim \mu_q \\ S \supseteq A}}[\tilde{g}_S(A) = 0] \leq
 \Pr_{\substack{S \sim \mu_q \\ S \supseteq A}}[\tilde{g}_S(A) = \tilde{g}(A)] \leq
 \Pr_{\substack{S \sim \mu_q \\ S \supseteq A}}[\tilde{g}_S(A) \neq 0],
\]
implying that
\[
 \Pr_{\substack{S \sim \mu_q \\ S \supseteq A}}[\tilde{g}_S(A) \neq 0] \geq \frac{1}{2},
\]
and so
\[
 \Pr_{S \sim \mu_q}[S \supseteq A \text{ and } \tilde{g}_S(A) \neq 0] \geq \frac{1}{2} q^{|A|}.
\]

Given $|R| \leq d$, we deduce that
\[
  \sum_{\substack{A \in \supp(g) \\ A \supseteq R}} \frac{1}{2} q^{|A|} \leq
 \sum_{\substack{A \in \supp(g) \\ A \supseteq R}} \Pr_{S \sim \mu_q}[S \supseteq A \text{ and } \tilde{g}_S(A) \neq 0] \leq
 \EE_{S \sim \mu_q}[|\supp(g_S)| \cdot \bbone_{S \supseteq R}] \leq q^{|R|} K.
\]
We conclude that the number of sets $A \in \supp(g)$ containing $R$ of size $|R| + e$ is at most $2K q^{-e} \leq (2K/q)^e$, as needed. 
\end{proof}

\subsection{Proof of \texorpdfstring{\Cref{lem:expansion}}{\ref{lem:expansion}}} \label{sec:lem:expansion}
It remains to prove \Cref{lem:expansion}. We first observe that we can sample $S_1,S_2$ on the left by sampling $S'_1,S'_2 \sim \mu_q(\{0,1\}^{\overline{A}})$ and then taking $S_b = S'_b \cup A$. We can similarly sample $S_1,S_2,T$ on the right via $(S'_1,S'_2,T') \sim \mu_{q,cq}(\{0,1\}^{\overline{A}})$. Henceforth we will omit the domain $\{0,1\}^{\overline{A}}$ in such expressions to avoid clutter.

For every $\sigma \in \mathbb{R}$ we define the indicator function $h_\sigma\colon \{0,1\}^{\overline{A}} \to \{0,1\}$ by
\[
 h_\sigma(S') = \indicator[\tilde{g}_{S' \cup A}(A) = \sigma].
\]
We can rephrase \Cref{lem:expansion} in these terms as follows:
\[
 \sum_\sigma \EE_{S'_1,S'_2 \sim \mu_q}[h_\sigma(S'_1) (1-h_\sigma(S'_2))] \leq \expT_c
 \sum_\sigma \EE_{(S'_1,S'_2,T) \sim \mu_{q,cq}}[h_\sigma(S'_1) (1-h_\sigma(S'_2))].
\]
\Cref{lem:expansion} thus reduces to the following statement.

\begin{lemma} \label{lem:expansion-aux}
Let $h\colon \{0,1\}^{n'} \to \{0,1\}$ be an indicator function. There exists a constant $\expT_c$, depending only on $c$, such that
\[
 \EE_{S'_1,S'_2 \sim \mu_q}[h(S'_1) (1 - h(S'_2))] \leq \expT_c
 \EE_{(S'_1,S'_2,T) \sim \mu_{q,cq}}[h(S'_1) (1 - h(S'_2))].
\]
\end{lemma}

At this point, we invoke Fourier analysis on $(\{0,1\}^{n'}, \mu_q)$. The Fourier basis for this domain is given by the functions
\[
 \omega_U(y) = \prod_{i \in U} \omega_i(y), \text{ where } \omega_i(y) = \frac{y_i - q}{\sqrt{q(1-q)}}.
\]
This basis is orthonormal with respect to $\mu_q$, and the corresponding Fourier expansion of $h$ is
\[
 h(y) = \sum_U \hat{h}(U) \omega_U(y),
\]
where $\hat{h}(\emptyset) = \EE[h]$. Since $h$ is Boolean, we furthermore have
\[
 \EE[h] = \EE[h^2] = \sum_U \hat{h}(U)^2.
\]
This allows us to express the left-hand side of \Cref{lem:expansion-aux} as
\[
 \EE_{S'_1,S'_2 \sim \mu_q}[h(S'_1) (1 - h(S'_2))] = \EE[h] - \EE[h]^2 = \sum_{U \neq \emptyset} \hat{h}(U)^2.
\]

In order to express the right-hand side in a similar form, it suffices to compute
\[
 \rho_c = \EE_{(S_1,S_2,T) \sim \mu_{q,cq}}[\omega_i(\bbone_{S_1}) \omega_i(\bbone_{S_2})].
\]
Indeed, since the marginal distributions of $S_1$ and $S_2$ are $\mu_q$, then
\[
 \EE_{(S_1,S_2,T) \sim \mu_{q,cq}}[\omega_i(\bbone_{S_1})] = \EE_{(S_1,S_2,T) \sim \mu_{q,cq}}[\omega_i(\bbone_{S_2})] = 0.
\]
The independence of coordinates implies that
\[
 \EE_{(S_1,S_2,T) \sim \mu_{q,cq}}[\omega_{U_1}(\bbone_{S_1}) \omega_{U_2}(\bbone_{S_2})] =
 \begin{cases}
     \rho_c^{|U_1|} & \text{if } U_1 = U_2, \\
     0 & \text{otherwise}.
 \end{cases}
\]
Therefore the expectation on the right-hand side of \Cref{lem:expansion-aux} is
\[
 \EE_{(S'_1,S'_2,T) \sim \mu_{q,cq}}[h(S'_1) (1 - h(S'_2))] =
 \EE[h] - \sum_U \rho_c^{|U|} \hat{h}(U)^2 =
 \sum_U (1 - \rho_c^{|U|}) \hat{h}(U)^2 \geq (1 - |\rho_c|) \sum_{U \neq \emptyset} \hat{h}(U)^2,
\]
since $|\rho_c| \leq 1$ by Cauchy--Schwarz.
It follows that \Cref{lem:expansion-aux} holds as long as $\expT_c \geq 1/(1 - \rho_c)$.

In order to complete the proof of \Cref{lem:expansion-aux}, we compute $\rho_c$. If $(S_1,S_2,T) \sim \mu_{q,cq}$ then $\Pr[(\bbone_{S_1})_i=1] = \Pr[(\bbone_{S_2})_i=1] = q$ and
\[
 \Pr[(\bbone_{S_1})_i=(\bbone_{S_2})_i=1] = cq + (1-cq) \frac{(q-cq)^2}{(1-cq)^2} = \frac{q^2 - 2cq^2 + cq}{1-cq}.
\]
Therefore
\[
 \rho_c =  \EE_{(S_1,S_2,T) \sim \mu_{q,cq}}\left[ \frac{((\bbone_{S_1})_i - q)((\bbone_{S_2})_i - q)}{q(1-q)}\right] =
 \frac{1}{q (1-q)} \left[\frac{q^2 - 2cq^2 + cq}{1-cq} - q^2 \right] =
 \frac{cq (1-q)}{(1-cq) q} \leq c,
\]
implying that we can take $\expT_c = 1/(1-c)$.

\section{Structure theorem}
\label{sec:structure}

In this section we prove our main structure theorem, \Cref{thm:main-intro}, which we reformulate using terminology defined in \Cref{sec:prel}.

\begin{theorem} \label{thm:structure}
Suppose that $f\colon (\{0,1\}^n, \mu_p) \to \{0,1\}$ is $\epsilon$-close to degree~$d$, where $p \leq 1/2$. Then $\Pr[f \neq g] = O(\epsilon)$ and $\EE[(f - g)^2] = O(\epsilon)$ for some function $g\colon (\{0,1\}^n, \mu_p) \to \mathbb{R}$ satisfying the following properties:
\begin{enumerate}[(i)]
\item If $y \in \{0,1\}^n$ has weight at most $d$ then $g(y) \in \{0,1\}$.
\label{itm:structure-boolean}
\item $g$ is a $(d, O_d(1/p))$-sparse junta.
\label{itm:structure-sparse}
\item For every $e$, the number of minimal non-Boolean inputs of $g$ of weight $e$ is $O_d(\epsilon/p^e)$.
\label{itm:structure-non-boolean}
\end{enumerate}
Conversely, if $g$ satisfies these properties, then it is $O_d(\epsilon)$-close to Boolean.
\end{theorem}

\begin{corollary} \label{cor:structure}
Let $f$ be as in the theorem. Then $f$ is $O_d(\epsilon)$-close to $\round(g, \{0,1\})$, where $g$ is the function in the theorem.

Conversely, if $g$ satisfies the conditions in the theorem, then $\round(g, \{0,1\})$ is $O_d(\epsilon)$-close to degree~$d$.
\end{corollary}
\begin{proof}[Proof of Corollary \ref{cor:structure}]
Let $f$ be as in the theorem. Then
\[
 \Pr[f \neq \round(g, \{0,1\})] \leq \Pr[f \neq g] = O(\epsilon).
\]

Conversely, suppose that $g$ satisfies the conditions in the theorem. Then $g$ has degree~$d$ and
\[
 \EE[(\round(g, \{0,1\}) - g)^2] = O_d(\epsilon)
\]
since $g$ is $O_d(\epsilon)$-close to Boolean.
\end{proof}

Henceforth, all big O constants will depend on $d$, and we do not mention this explicitly.

Most of the effort will be concentrated on proving the direct part of the theorem, which we will do in two steps:
\begin{enumerate}
    \item We first construct a sparse junta $h$ which is $O(\epsilon)$-close to $f$.
    \item We then modify $h$ to a sparse junta $g$ which satisfies the first two properties in the theorem, and show that it satisfies the third property as well.
\end{enumerate}
We then prove the easier converse part of the theorem.

\subsection{Step 1: Agreement}
\label{sec:structure-agreement}

Since $f$ is $\epsilon$-close to degree~$d$, there is a degree~$d$ function $f^{\leq d}$ such that $\EE_{\mu_p}[(f - f^{\leq d})^2] \leq \epsilon$.

For $S \subseteq [n]$, define
\[
 \epsilon_S = \EE_{\mu_{1/2}}[(f|_S - f^{\leq d}|_S)^2],
\]
where $\mu_{1/2}$ refers to the distribution $\mu_{1/2}(\{0,1\}^S)$; below we usually omit the domain.
Since sampling $S \sim \mu_{2p}$ and $y \sim \mu_{1/2}(\{0,1\}^S)$ is the same as sampling $y \sim \mu_p$,
\[
 \EE_{S \sim \mu_{2p}}[\epsilon_S] = \EE_{\mu_p}[(f - f^{\leq d})^2] \leq \epsilon.
\]
Applying the Kindler--Safra theorem (\Cref{thm:kindler-safra}), there exist Boolean degree~$d$ functions $g_S$ such that $\EE_{\mu_{1/2}}[(f|_S - g_S)^2] = O(\epsilon_S)$,
and so
\[
 \EE_{S \sim \mu_{2p}} \left[ \EE_{\mu_{1/2}}[(f|_S - g_S)^2] \right] =
 O\left( \EE_{S \sim \mu_{2p}}[\epsilon_S] \right) = 
 O(\epsilon).
\]

We would like to paste the functions $g_S$ to a global function $h$ using the junta agreement theorem (\Cref{thm:agreement}), which we will apply with $K = M_d$ (the value from \Cref{thm:nisan-szegedy}), $c = \sqrt{1/2}$, and $q = 2p$. \Cref{thm:nisan-szegedy} shows that each $g_S$ is a $K$-junta. In order to apply the junta agreement theorem, we need to verify the agreement condition:
\[
 \Pr_{(S_1,S_2,T) \sim \mu_{2p,\sqrt{2}p}}[g_{S_1}|_T \neq g_{S_2}|_T] = O(\epsilon).
\]

To see this, first notice that since $g_{S_1}$ and $g_{S_2}$ are both $K$-juntas, if $g_{S_1}|_T \neq g_{S_2}|_T$ then $\Pr_{y \sim \mu_{\sqrt{1/2}}}[g_{S_1}|_T(y) \neq g_{S_2}|_T(y)] \geq \sqrt{1/2}^{2K} = 2^{-K}$. Since $g_{S_1}$ and $g_{S_2}$ are Boolean, this shows that
\[
 \Pr_{(S_1,S_2,T) \sim \mu_{2p,\sqrt{2}p}}[g_{S_1}|_T \neq g_{S_2}|_T] \leq
 2^K \EE_{(S_1,S_2,T) \sim \mu_{2p,\sqrt{2}p}} \left[
 \EE_{\mu_{\sqrt{1/2}}}[(g_{S_1}|_T - g_{S_2}|_T)^2]
 \right].
\]
Applying the inequality $(g_{S_1}|_T - g_{S_2}|_T)^2 \leq 2(g_{S_1}|_T - f|_T)^2 + 2(g_{S_2}|_T - f|_T)^2$, we obtain
\[
 \Pr_{(S_1,S_2,T) \sim \mu_{2p,\sqrt{2}p}}[g_{S_1}|_T \neq g_{S_2}|_T] \leq
 2^{K+1} \EE_{(S_1,S_2,T) \sim \mu_{2p,\sqrt{2}p}}  \left[
 \EE_{\mu_{\sqrt{1/2}}}[(g_{S_1}|_T - f|_T)^2]
 \right].
\]
We can choose $(S_1,T)$ in the following way: choose $S_1 \sim \mu_{2p}$, and choose $T \sim \mu_{\sqrt{1/2}}(S_1)$. If we then choose $y \sim \mu_{\sqrt{1/2}}(T)$ then it is equivalent to directly sampling $y \sim \mu_{1/2}(S_1)$.
This shows that
\[
 \Pr_{(S_1,S_2,T) \sim \mu_{2p,\sqrt{2}p}}[g_{S_1}|_T \neq g_{S_2}|_T] \leq
 2^{K+1} \EE_{S \sim \mu_{2p}}\left[ \EE_{\mu_{1/2}}[(g_S - f|_S)^2]\right] = 
 O(\epsilon),
\]
and so the agreement condition is satisfied.

Applying the junta agreement theorem, we obtain a $(d,O(1/p))$-sparse junta $h$ such that
\[
 \Pr_{S \sim \mu_{2p}}[h|_S \neq g_S] = O(\epsilon).
\]
Furthermore, each monomial coefficient of $h$ is a monomial coefficient of some Boolean degree~$d$ function, and so belongs to $\{-2^{d-1}, \dots, 2^{d-1}\}$ by \Cref{lem:granularity}.

\medskip

To conclude this step of the proof, we need to show that $\EE_{\mu_p}[(f - h)^2] = O(\epsilon)$. We first show that $\Pr_{\mu_p}[f \neq h] = O(\epsilon)$. Indeed,
\[
 \Pr_{\mu_p}[f \neq h] = \EE_{S \sim \mu_{2p}} \left[
 \Pr_{\mu_{1/2}} [f|_S \neq h|_S]
 \right] \leq
 \EE_{S \sim \mu_{2p}} \left[
 \Pr_{\mu_{1/2}} [f|_S \neq g_S] + \indicator[g_S \neq h|_S]
 \right] = O(\epsilon).
\]
(The term $\indicator[g_S \neq h|_S]$ is an indicator.)

In order to convert this $L_0$ guarantee to an $L_2$ guarantee, we use \Cref{lem:L0-L2}, which implies that
\[
 \EE_{\mu_p}[\dist(h, \{0,1\})^2] = O(\Pr_{\mu_p}[h \notin \{0,1\}]) =
 O(\Pr_{\mu_p}[h \neq f]) = O(\epsilon).
\]
This implies that
\[
 \EE_{\mu_p}[(f - h)^2] \leq
 2\EE_{\mu_p}[(f - \round(h,\{0,1\}))^2] +
 2\EE_{\mu_p}[(\round(h,\{0,1\}) - h)^2] =
 2\Pr_{\mu_p}[f \neq h] + O(\epsilon) = O(\epsilon).
\]

Concluding, in this step we have constructed a function $h$ satisfying the following properties:
\begin{enumerate}[(i)]
    \item $h$ is a $(d,C/p)$-sparse junta.
    \item $h$ is $B$-quantized.
    \item $\Pr[h \neq f] = O(\epsilon)$.
    \item $\EE[(h - f)^2] = O(\epsilon)$.
\end{enumerate}
In the final two properties, the underlying distribution is $\mu_p$, which we assume henceforth.

\subsection{Step 2: Fixing coefficients}
\label{sec:structure-fixing}

In this step we modify $h$ to a function $g$ which satisfies all properties listed in \Cref{thm:structure}. We do this by constructing a sequence of functions $h = h_{-1}, h_0, \dots, h_d = g$, where $h_e$ satisfies the following properties:
\begin{enumerate}[(i)]
    \item $h_e$ is a $(d,C_e/p)$-sparse junta, where $C_e$ depends only on $d,e$.
    \label{itm:step2-sparse}
    \item $h_e$ is $B_e$-quantized for some finite $B_e$ depending only on $d,e$.
    \label{itm:step2-quantized}
    \item $\Pr[h_e \neq f] = O_{d,e}(\epsilon)$.
    \label{itm:step2-L0}
    \item $\EE[(h_e - f)^2] = O_{d,e}(\epsilon)$.
    \label{itm:step2-L2}
    \item $h_e(y) \in \{0,1\}$ if $y$ has weight at most $e$.
    \label{itm:step2-boolean}
\end{enumerate}

The function $h$ clearly satisfies the properties for $e = -1$. We now show how to construct $h_e$ from $h_{e-1}$ for $e \in \{0,\ldots,d\}$.

Given $h_{e-1}$, let $\mathcal{F}$ consist of all minimal non-Boolean inputs of $h_{e-1}$ of weight $e$. We define $h_e$ by modifying $\tilde{h}_e(S)$ for each $\bbone_S \in \mathcal{F}$ so that $h_e(\bbone_S) = 0$:
\[
 \tilde{h}_e(S) = -\sum_{T \subsetneq S} \tilde{h}_{e-1}(T).
\]
All other monomial coefficients stay the same.

\paragraph{\Cref{itm:step2-sparse}.} \Cref{lem:minimal-non-boolean} shows that $\mathcal{F} \subseteq \supp(h_{e-1})^{\ssum L_d}$. \Cref{lem:sparsity-elementary} shows that $\mathcal{G} = \supp(h_{e-1})^{\ssum L_d}$ is $(L_d d,C'_e/p)$-sparse, for some constant $C'_e$. Since $\supp(h_e) \subseteq \supp(h_{e-1}) \cup \mathcal{F}$, \Cref{lem:sparsity-elementary} shows that $h_e$ is a $(d,C_e/p)$-sparse junta, for some constant $C_e$.

\paragraph{\Cref{itm:step2-quantized}.} Each modified monomial coefficient is the negated sum of $2^e - 1$ monomial coefficients of $h_{e-1}$, and so $h_e$ is $B_e$-quantized for some finite set $B_e$.

\paragraph{\Cref{itm:step2-L0}.}
If $h_e(y) \neq h_{e-1}(y)$ then $y_S = 1$ for some $S \in \mathcal{F}$, and so
\[
 \Pr[h_e \neq h_{e-1}] \leq \sum_{S \in \mathcal{F}} p^{|S|}.
\]
In order to bound this, we apply the reverse union bound (\Cref{lem:reverse-union-bound}),
concluding that
\[
 \Pr[h_e \neq h_{e-1}] = O(\Pr[\mathcal{E}]),
\]
where $\mathcal{E}$ is the event that for some $S \in \mathcal{F}$, it holds that $y_S = 1$ and $y_R = 0$ for all $R \in \mathcal{G}$ such that $R \not\subseteq S$. If $\mathcal{E}$ happens then $h_{e-1}(y) = h_{e-1}(\bbone_S) \notin \{0,1\}$, and so
\[
 \Pr[h_e \neq h_{e-1}] = O(\Pr[h_{e-1} \notin \{0,1\}]) = O(\Pr[h_{e-1} \neq f]) = O(\epsilon).
\]
Therefore
\[
 \Pr[h_e \neq f] \leq \Pr[h_e \neq h_{e-1}] + \Pr[h_{e-1} \neq f] = O(\epsilon).
\]

\paragraph{\Cref{itm:step2-L2}.} \Cref{lem:sparsity-elementary,lem:quantized} show that $h_e - h_{e-1}$ is a $B^+_e$-quantized $(d,O(1/p))$-sparse junta, for some finite set $B^+_e$. Therefore \Cref{lem:L0-L2-norm}, applied with $f = h_e - h_{e-1}$, shows that
\[
 \EE[(h_{e-1} - h_e)^2] = O(\Pr[h_{e-1} \neq h_e]) = O(\epsilon).
\]
Consequently
\[
 \EE[(h_e - f)^2] \leq 2\EE[(h_e - h_{e-1})^2] + 2\EE[(h_{e-1} - f)^2] = O(\epsilon).
\]

\paragraph{\Cref{itm:step2-boolean}.} Since we only modified monomial coefficients of size $e$, clearly $h_e(y) \in \{0,1\}$ for $y$ of weight less than $e$. By construction, $h_e(y) \in \{0,1\}$ for $y$ of weight exactly $e$.

\subsubsection*{Concluding the proof} Taking $g = h_d$, the function $g$ satisfies all properties stated in \Cref{thm:structure}, except for \Cref{itm:structure-non-boolean}, which states that the number of minimal non-Boolean inputs of $g$ of weight $e$ is $O(\epsilon/p^e)$.

Let $\mathcal{F}$ consist of all minimal non-Boolean inputs of $g$. \Cref{lem:minimal-non-boolean} shows that $\mathcal{F} \subseteq \mathcal{G} = \supp(g)^{\ssum L_d}$. Applying \Cref{lem:sparsity-elementary} and the reverse union bound just as above, we conclude that
\[
 \sum_{S \in \mathcal{F}} p^{|S|} = O(\Pr[g \notin \{0,1\}]) = O(\Pr[g \neq f]) = O(\epsilon).
\]
Therefore $\mathcal{F}$ contains at most $O(\epsilon/p^e)$ inputs of weight $e$.

\subsection{Converse part}
\label{sec:structure-converse}

We conclude the proof of \Cref{thm:structure} by proving the converse direction. Suppose that $g$ satisfies the three properties in the statement of the theorem:
\begin{enumerate}[(i)]
\item $g(y) \in \{0,1\}$ whenever $y \in \{0,1\}^n$ has weight at most $d$.
\item $g$ is a $(d,O(1/p))$-sparse junta.
\item $g$ has $O(\epsilon/p^e)$ minimal non-Boolean inputs of weight $e$, for every $e$.
\end{enumerate}
We need to show that $g$ is $O(\epsilon)$-close to Boolean.

We start by showing that the monomial coefficients of $g$ are quantized.
If $|A| \leq d$ then $g|_A$ is a Boolean function by \Cref{itm:structure-boolean}, and so $\tilde{g}(A) \in \{-2^{d-1},\dots,2^{d-1}\}$ by \Cref{lem:granularity}.

If $g(y) \notin \{0,1\}$ then $y \geq z$ for some minimal non-Boolean input $z$. If $z$ has weight $e$ then the probability that $y \geq z$ is $p^e$. \Cref{lem:minimal-non-boolean} shows that $z$ has weight at most $L_d$, and so
\[
 \Pr[g \notin \{0,1\}] \leq \sum_{e \leq L_d} p^e \cdot O(\epsilon/p^e) = O(\epsilon).
\]
\Cref{lem:L0-L2} concludes that $g$ is $O(\epsilon)$-close to Boolean.

\section{Monotone structure theorems}
\label{sec:structure-monotone}

In this section we prove \Cref{thm:main-monotone-intro}. We start by proving a monotone version of \Cref{thm:structure}, which is interesting in its own right, and then deduce \Cref{thm:main-monotone-intro} from it.

\subsection{Monotone version of \texorpdfstring{\Cref{thm:structure}}{\ref{thm:structure}}}
\label{sec:structure-monotone-version}

We start by proving a monotone version of \Cref{thm:structure}.

\begin{theorem} \label{thm:structure-monotone}
Suppose that $f\colon (\{0,1\}^n, \mu_p) \to \{0,1\}$ is monotone and $\epsilon$-close to degree~$d$, where $p \leq 1/2$. Then $\Pr[f \neq g] = O(\epsilon)$ and $\EE[(f - g)^2] = O(\epsilon)$ for some function $g\colon (\{0,1\}^n, \mu_p) \to \mathbb{R}$ satisfying the following properties:
\begin{enumerate}[(i)]
\item If $|S| \leq d$ then $g|_S$ is monotone and Boolean.
\label{itm:structure-monotone-boolean}
\item $g$ is a $(d,O_d(1/p))$-sparse junta.
\label{itm:structure-monotone-sparse}
\item For every $e$, the number of minimal non-Boolean inputs of $g$ of weight $e$ is $O_d(\epsilon/p^e)$.
\label{itm:structure-monotone-non-boolean}
\item For every $e$, the number of minimal non-monotone Boolean inputs of $g$ of weight $e$ is $O_d(\epsilon/p^e)$.
\label{itm:structure-monotone-non-monotone}
\end{enumerate}
Conversely, if $g$ satisfies these properties then it is $O_d(\epsilon)$-close to some monotone Boolean function.
\end{theorem}

\begin{corollary} \label{cor:structure-monotone}
Let $f$ be as in theorem. Then $f$ is $O_d(\epsilon)$-close to $\round(g, \{0,1\})$, where $g$ is the function in the theorem.

Conversely, if $g$ satisfies the conditions in the theorem, then $\round(g, \{0,1\})$ is $O_d(\epsilon)$-close to a monotone Boolean function which is $O_d(\epsilon)$-close to degree~$d$.
\end{corollary}
\begin{proof}[Proof of Corollary \ref{cor:structure-monotone}]
The first statement follows as in the proof of \Cref{cor:structure}. Now suppose that $g$ satisfies the conditions in the theorem. According to the theorem, there exists some monotone Boolean function $f$ such that $\EE[(g - f)^2] = O_d(\epsilon)$. Since $g$ has degree~$d$, $f$ is $O_d(\epsilon)$-close to degree~$d$.
\end{proof}

The proof of the direct part of \Cref{thm:structure-monotone} closely resembles the proof of the direct part of \Cref{thm:structure}, while the proof of the converse part is slightly more tricky. For both proofs, we suppress the dependence of big $O $constants on $d$.

\subsubsection*{Direct part}

Suppose that $f\colon (\{0,1\}^n, \mu_p) \to \{0,1\}$ is monotone and $\epsilon$-close to degree~$d$. The argument in \Cref{sec:structure-agreement} applies, constructing a function $h$ which satisfies the properties stated there (we repeat them below as part of the proof). We now run an argument in the style of \Cref{sec:structure-fixing}, constructing a sequence of functions $h = h_{-1},h_0,\ldots,h_d = g$, where $h_e$ satisfies the following properties:
\begin{enumerate}[(i)]
    \item $h_e$ is a $(d,C_e/p)$-sparse junta, where $C_e$ depends only on $d,e$.
    \label{itm:step2m-sparse}
    \item $h_e$ is $B_e$-quantized for some finite $B_e$ depending only on $d,e$.
    \label{itm:step2m-quantized}
    \item $\Pr[h_e \neq f] = O_{d,e}(\epsilon)$.
    \label{itm:step2m-L0}
    \item $\EE[(h_e - f)^2] = O_{d,e}(\epsilon)$.
    \label{itm:step2m-L2}
    \item $h_e|_S$ is a monotone Boolean function if $|S| \leq e$.
    \label{itm:step2m-boolean}
\end{enumerate}

The function $h$ constructed in \Cref{sec:structure-agreement} satisfies these properties for $e = -1$. We now show how to construct $h_e$ from $h_{e-1}$ for $e \in \{0,\ldots,d\}$, using a modification of the construction in \Cref{sec:structure-fixing}.

Given $h_{e-1}$, let $\mathcal{F}_b$ consist of all minimal non-Boolean inputs of $h_{e-1}$ of weight $e$, let $\mathcal{F}_m$ consist of all minimal non-monotone Boolean inputs of $h_{e-1}$ of weight $e$, and let $\mathcal{F} = \mathcal{F}_b \cup \mathcal{F}_m$.
We define $h_e$ by modifying $\tilde{h}_e(S)$ for each $\bbone_S \in \mathcal{F}$ so that $h_e(\bbone_S) = 1$. All other monomial coefficients stay the same.

\Cref{itm:step2m-sparse} follows just as in \Cref{sec:structure-fixing}, using \Cref{lem:minimal-non-monotone} in addition to \Cref{lem:minimal-non-boolean}. We need to take $\mathcal{G} = \supp(f)^{\ssum \max(L_d,M_d)}$.
\Cref{itm:step2m-quantized} follows as in \Cref{sec:structure-fixing}.

\Cref{itm:step2m-L0} requires a new argument. If $h_e(y) \neq h_{e-1}(y)$ then $y_S = 1$ for some $S \in \mathcal{F}$, and so
\[
 \Pr[h_e \neq h_{e-1}] \leq \sum_{S \in \mathcal{F}_b} p^{|S|} + \sum_{S \in \mathcal{F}_m} p^{|S|}.
\]

The argument in \Cref{sec:structure-fixing} shows that the first term on the right is $O(\epsilon)$. In order to bound the second term, we apply the reverse union bound (\Cref{lem:reverse-union-bound}) with $p := 2p$, concluding that
\[
 \sum_{S \in \mathcal{F}_m} p^{|S|} \leq
 \sum_{S \in \mathcal{F}_m} (2p)^{|S|} = O(\Pr_{\mu_{2p}}[\mathcal{E}]),
\]
where $\mathcal{E}$ is the event that for some $S \in \mathcal{F}$, it holds that $y_S = 1$ and $y_R = 0$ for all $R \in \supp(\mathcal{G})$ such that $R \not\subseteq S$.

If $\mathcal{E}$ happens for some $S$, an event we denote by $\mathcal{E}_S$, then $h_{e-1}|_S$ is a non-monotone Boolean function of degree~$d$, which is an $M_d$-junta by \Cref{thm:nisan-szegedy}. Therefore \Cref{lem:non-monotone-junta} implies that $\Pr_{\mu_{1/2}}[h_{e-1}|_S \neq f|_S] \geq 2^{-M_d}$. Thus
\[
 \Pr_{\mu_p}[h_{e-1} \neq f] =
 \Pr_{\substack{S \sim \mu_{2p} \\ y \sim \mu_{1/2}(S)}}[h(y) \neq f(y)] \geq 2^{-M_d} \Pr_{\mu_{2p}}[\mathcal{E}].
\]
Since the left-hand side is $O(\epsilon)$, we conclude that
\[
 \sum_{S \in \mathcal{F}_m} p^{|S|} = O(\Pr_{\mu_{2p}}[\mathcal{E}]) = O(\epsilon).
\]
This shows that $\Pr[h_e \neq h_{e-1}] = O(\epsilon)$, and so $\Pr[h_e \neq f] \leq \Pr[h_e \neq h_{e-1}] + \Pr[h_{e-1} \neq f] = O(\epsilon)$.

\Cref{itm:step2m-L2} follows from \Cref{itm:step2m-L0} as in \Cref{sec:structure-fixing}. \Cref{itm:step2m-boolean} follows by construction.

\paragraph{Concluding the proof} Taking $g = h_d$, the function $g$ satisfies all properties stated in \Cref{thm:structure-monotone} except for \Cref{itm:structure-monotone-non-boolean,itm:structure-monotone-non-monotone}.

\Cref{itm:structure-monotone-non-boolean} follows as in \Cref{sec:structure-fixing}. In order to prove \Cref{itm:structure-monotone-non-monotone}, let $\mathcal{F}_m$ denote all minimal non-monotone Boolean inputs of $g$. The proof of \Cref{itm:step2m-L0} above shows that
\[
 \sum_{S \in \mathcal{F}_m} p^{|S|} = O(\epsilon),
\]
from which \Cref{itm:structure-monotone-non-monotone} of the theorem immediately follows.

\subsubsection*{Converse part}

We conclude the proof of \Cref{thm:structure-monotone} by proving the converse direction. Suppose that $g$ satisfies the properties in the statement of the theorem:
\begin{enumerate}[(i)]
\item $g(y) \in \{0,1\}$ whenever $y \in \{0,1\}^n$ has weight at most $d$.
\item $g$ is a $(d,O(1/p))$-sparse junta.
\item $g$ has $O(\epsilon/p^e)$ minimal non-Boolean inputs of weight $e$, for all $e$.
\item $g$ has $O(\epsilon/p^e)$ minimal non-monotone Boolean inputs of weight $e$, for all $e$.
\end{enumerate}
We need to show that $g$ is $O(\epsilon)$-close to some monotone Boolean function $f$.

\Cref{thm:structure} shows that $g$ is $O(\epsilon)$-close to the Boolean function $G = \round(g, \{0,1\})$. However, $G$ is not necessarily monotone. Accordingly, we define
\[
 f(y) = \max_{z \leq y} G(z),
\]
which is monotone by construction. It remains to show that $g$ is $O(\epsilon)$-close to $f$. We will show that $G$ is $O(\epsilon)$-close to $f$. This would imply that
\[
 \EE[(g - f)^2] \leq 2\EE[(g - G)^2] + 2\EE[(G - f)^2] = O(\epsilon) + 2\Pr[G \neq f] = O(\epsilon).
\]

If $y = \bbone_S \in \{0,1\}^n$ is such that no $z \leq y$ is a minimal non-Boolean input or a minimal non-monotone Boolean input of $g$ then $G|_S = g|_S$ and $g|_S$ is monotone, implying that $f(y) = G(y)$. Therefore if $f(y) \neq G(y)$ then $y \ge z$ for some input $z = \bbone_T$ which is either a minimal non-Boolean input of $g$ or a minimal non-monotone Boolean input of $g$. Since $y \geq z$ with probability~$p^{|T|}$, if we let $\mathcal{F}$ consist of all minimal non-Boolean and minimal non-monotone Boolean inputs of $g$ then
\[
 \Pr[G \neq f] \leq \sum_{\bbone_T \in \mathcal{F}} p^{|T|} \leq \sum_{e \leq \max(L_d,M_d)} p^e \cdot O(\epsilon/p^e) = O(\epsilon),
\]
using \Cref{lem:minimal-non-boolean,lem:minimal-non-monotone} to bound the size of an input in $\mathcal{F}$. This concludes the proof.

\subsection{Approximation by monotone DNF}
\label{sec:structure-monotone-DNF}

We now deduce \Cref{thm:main-monotone-intro}, which we rephrase as follows.

\begin{theorem} \label{thm:structure-monotone-DNF}
Suppose that $f\colon (\{0,1\}^n, \mu_p) \to \{0,1\}$ is monotone and $\epsilon$-close to degree~$d$, where $p \leq 1/2$. Then $f$ is $O(\epsilon)$-close to a monotone DNF $F$ which satisfies the following properties:
\begin{enumerate}[(i)]
\item $F$ is a $(d,O_d(1/p))$-sparse DNF.
\label{itm:structure-DNF-sparse}
\item For every $e$, the number of minimal high-degree inputs of $F$ of weight $e$ is $O_d(\epsilon/p^e)$.
\label{itm:structure-DNF-high-degree}
\end{enumerate}
Conversely, if $F$ is a monotone DNF satisfying these properties then it is $O_d(\epsilon)$-close to degree~$d$.
\end{theorem}

We derive \Cref{thm:structure-monotone-DNF} from \Cref{thm:structure-monotone}. In the proof, we suppress the dependence of the big $O$ constants on $d$.

\subsubsection*{Direct part}

Suppose that $f\colon (\{0,1\}^n, \mu_p) \to \{0,1\}$ is monotone and $\epsilon$-close to degree~$d$. Applying the direct part of \Cref{thm:structure-monotone}, we obtain a function $g$ satisfying the stated properties which is $O(\epsilon)$-close to $f$. Applying the converse part of \Cref{thm:structure-monotone}, we obtain a monotone Boolean function $F$ which is $O(\epsilon)$-close to $g$, and so $O(\epsilon)$-close to $f$.

We first show that all minterms of $F$ have weight at most~$d$, using the way that $F$ is constructed in the proof of the converse part of \Cref{thm:structure-monotone}.
Recall that $G(y) = \round(g(y), \{0,1\})$. Suppose that $\bbone_S$ is an input of weight larger than $d$ such that $F(z) = 0$ for all $z < \bbone_S$, and so $G(z) = 0$ for all $z < \bbone_S$. If $z = \bbone_T$ has weight at most $d$ then $g|_T = G|_T = 0$, and so $\tilde{g}(A) = 0$ for all $A \subseteq T$. It follows that $\tilde{g}(A) = 0$ for all $A \subseteq S$, and so $G(z) = 0$ for all $z \leq \bbone_S$, implying that $F(\bbone_S) = 0$. Thus $\bbone_S$ cannot be a minterm of $F$.

Thus, $F$ is a monotone width~$d$ DNF, whose minterms are all $y$ of weight at most $d$ such that $g(y) = 1$ and $g(z) = 0$ for all $z < y$. If $y = \bbone_S$ is such a minterm then $g|_S = y_S$, and so $\tilde{g}(S) = 1$. Therefore the minterm support of $F$ is contained in the monomial support of $g$. Since $g$ is $(d,O(1/p))$-sparse, it follows that $F$ is also $(d,O(1/p))$-sparse, proving \Cref{itm:structure-DNF-sparse}.

\smallskip

It remains to prove \Cref{itm:structure-DNF-high-degree}. We first show that $F|_S = g|_S$ if and only if $F|_S$ has degree~$d$.
If $F|_S = g|_S$ then $F|_S$ has degree~$d$. Conversely, suppose that $F|_S$ has degree~$d$. Since $g|_A$ is Boolean and monotone whenever $|A| \leq d$, we see that $F|_A = g|_A$ and so $\tilde{F}(A) = \tilde{g}(A)$ for all $A \subseteq S$ of size at most~$d$. Since $F|_S$ has degree $d$, this means that $F|_S = g|_S$.

We conclude that if $y$ is a minimal high-degree input of $F$ then it is a minimal input such that $F(y) \neq g(y)$. In particular, $g|_T = F|_T$ is Boolean and monotone for all $\bbone_T < y$.
Since $F(y) \neq g(y)$, either $g(y) \neq G(y)$ or $g(y) = G(y) \neq F(y)$. If $g(y) \neq G(y)$ then $g(y) \notin \{0,1\}$, and so $y$ is a minimal non-Boolean input of $g$. If $g(y) = G(y) \neq F(y)$ then $y$ is a minimal non-monotone Boolean input of $g$. \Cref{itm:structure-DNF-high-degree} now follows from \Cref{itm:structure-monotone-non-boolean,itm:structure-monotone-non-monotone} of \Cref{thm:structure-monotone}.

\subsubsection*{Converse part}

Let $F$ be a monotone DNF satisfying the properties listed in \Cref{thm:structure-monotone-DNF}. Define
\[
 g = \sum_{|A| \leq d} \tilde{F}(A) y_A.
\]

Observe that $g|_S = F|_S$ if and only if $F|_S$ has degree~$d$. In particular, if $g(y) \neq F(y)$ then $y \geq z$ for some minimal high-degree input $z$ of $g$. Denoting by $\mathcal{H}$ the set of all minimal high-degree inputs of $g$, this implies that
\[
 \Pr[g \neq F] \leq \sum_{\bbone_T \in \mathcal{H}}
 p^{|T|} \leq \sum_{e \leq L_d} p^e \cdot O(\epsilon/p^e) = O(\epsilon),
\]
using \Cref{lem:minimal-high-degree} to bound the weight of a minimal high-degree input.

In order to upgrade this to an $L_2$ bound, we appeal to \Cref{lem:L0-L2}. First, we need to show that $g$ is a quantized sparse junta.

Since every monomial coefficient of $g$ is a monomial coefficient of some (monotone) Boolean function on $d$ coordinates, $g$ is $B$-quantized for some finite $B$.

In order to show that $g$ is a sparse junta, let $\mathcal{M}$ be the set of minterms of $F$, and observe that
\[
 F = \sum_e (-1)^e \sum_{S_1,\ldots,S_e \in \mathcal{M}} y_{S_1 \cup \cdots \cup S_e}.
\]
In particular, $\supp(g) \subseteq \supp(F)^{\ssum d}$, where $\supp(F)$ is the monomial support of $F$. Since $F$ is a $(d,O(1/p))$-sparse DNF, \Cref{lem:sparsity-elementary} implies that $g$ is a $(d,O(1/p))$-sparse junta.

Applying \Cref{lem:L0-L2}, we see that
\[
 \EE[\dist(g, \{0,1\})^2] = O(\Pr[g \notin \{0,1\}]) = O(\Pr[g \neq F]) = O(\epsilon).
\]
Therefore
\[
 \EE[(g - F)^2] \leq 2\EE[(g - \round(g,\{0,1\})^2] + 2\EE[(\round(g,\{0,1\}) - F)^2] \leq O(\epsilon) + 2\Pr[g \neq F] = O(\epsilon).
\]

\section{Junta approximation}
\label{sec:junta}

In this section we prove \Cref{thm:junta-intro,thm:junta-intro-formula}. Before restating them, we formally define the parameter appearing in them.

\begin{definition}
\label{def:junta-parameter}
 Given $d \in \mathbb{N}$, the parameter $m_d$ is the maximum $m$ such that there exists a degree~$d$ polynomial $P$ satisfying $P(0),\ldots,P(m-1) \in \{0,1\}$ and $P(0) \neq P(1)$.
\end{definition}

It might not be clear from the definition that $m_d < \infty$. In fact, $m_d \leq 2d$. To see this, suppose that $P$ is a degree~$d$ polynomial $P$ such that $P(0),\ldots,P(m-1) \in \{0,1\}$ and $P(0) \neq P(1)$. Since $P(0) \neq P(1)$, the polynomial $P$ is not constant, and so there can be at most $d$ values $x$ such that $P(x) = 0$ and at most $d$ values $y$ such that $P(y) = 1$. Consequently $m \leq 2d$. In contrast, the polynomial $P(x) = (1-x)\cdots(d-x)/d!$ shows that $m_d \geq d+1$.

Von zur Gathen and Roche~\cite{GathenR1997} considered a similar parameter $m'_d$ in which the condition $P(0) \neq P(1)$ is replaced by the condition that $P$ be non-constant. They obtained the improved bound $m'_d \leq d + O(d^{0.525})$, and conjectured that in fact $m'_d \leq d + O(1)$. Since $m_d \leq m'_d$, their bound also holds for $m_d$.

We can now restate the two theorems from the introduction, as a single theorem.

\begin{theorem} \label{thm:junta}
Suppose that $f\colon (\{0,1\}^n, \mu_p) \to \{0,1\}$ is $\epsilon$-close to degree~$d$, where $p \leq 1/2$.
Then for some constant $\mexp_d$ depending only on $d$,
\begin{enumerate}[(a)]
\item $f$ is $O_d(\epsilon^{1/m_d} + p)$-close to a constant Boolean function.
\label{itm:junta-constant}
\item $f$ is $O_d(\epsilon^{1/\mexp_d})$-close to a Boolean degree~$d$ function.
\label{itm:junta-boolean-degree-d}
\end{enumerate}
Furthermore, the first item becomes false if we replace $m_d$ by any smaller constant (for any value of the constant hidden in the big $O$).
\end{theorem}
In the rest of this section, we suppress the dependence of big $O$ constants on $d$.

\paragraph{The constant $\mexp_d$.}
We do not know the optimal value of $\mexp_d$, and in particular, whether it equals $m_d$.
If we consider \emph{$A$-valued} functions instead of Boolean functions, a setting we define below, then it is easy to construct examples in which $m_{d,A} \neq \mexp_{d,A}$, where $m_{d,A},\mexp_{d,A}$ are the corresponding parameters in the $A$-valued setting.

Let $A \subseteq \mathbb{R}$ be a finite set.
If $f(x) \in A$ for all $x$ in the domain of $f$ then we say that $f$ is $A$-valued. A $\{0,1\}$-valued function is thus the same as a Boolean function. Everything in this paper generalizes to the $A$-valued setting, once we generalize the Kindler--Safra theorem to the $A$-valued setting (this can be done in a black-box way, as shown in the full version of~\cite{DinurFH2019}, available on arXiv (1711.09428) and ECCC (TR17-180)).

When $d = 1$, it is not hard to check that $m_{1,A}$ is the size of the largest arithmetic progression in $A$. Therefore if $a,b,c$ are linearly independent over $\mathbb{Q}$ then the alphabet
\[
 A = \{0, a, b, c, a+b, a+c, b+c \}
\]
satisfies $m_{1,A} = 2$. Now consider the function
\[
 f = a y_1 + b y_2 + c y_3.
\]
By construction, $f$ is $\Theta(p^3)$-close to $A$, and so $F = \round(f, A)$ is $\Theta(p^3)$-close to degree~$1$. All $A$-valued degree~$1$ functions depend on at most two coordinates, and so are $\Omega(p)$-far from $f$. This shows that $\mexp_{d,A} \geq 3$.

\subsection{Converse part}
\label{sec:junta-converse}

The converse part of \Cref{thm:junta} is easy to prove, and serves to motivate the statement.

\begin{proof}[Proof of converse part]
Let $P$ be a degree~$d$ polynomial such that $P(0),\ldots,\linebreak[4]P(m_d-1) \in \{0,1\}$ and $P(0) \neq P(1)$. For any $p \leq 1/2$ and $\epsilon \leq 1$, define
\[
 f_{p,\epsilon} = P(y_1 + \cdots + y_N), \text{ where } N = \epsilon^{1/m_d}/p.
\]
We assume that $N$ is an integer, and in particular, $\epsilon^{1/m_d} \geq p$.

If fewer than $m_d$ of $y_1,\ldots,y_N$ evaluate to~$1$ then $f_{p,\epsilon} \in \{0,1\}$, and so
\[
 \Pr[f_{p,\epsilon} \notin \{0,1\}] \leq N^{m_d} p^{m_d} = \epsilon.
\]

In order to upgrade this to an $L_2$ bound, we use \Cref{lem:L0-L2}.
Since $y_1 + \cdots + y_N$ is a $(1,1/p)$-sparse junta, \Cref{lem:sparsity-elementary} shows that $f_{p,\epsilon}$ is a $(d,O(1/p))$-sparse junta. In order to show that $f_{p,\epsilon}$ is quantized, notice that we can find coefficients $c_e$ such that
\[
 P(n) = \sum_{e \leq d} c_e \binom{n}{e},
\]
and so
\[
 f_{p,\epsilon} = \sum_{e \leq d} c_e \sum_{\substack{S \subseteq [N] \\ |S| = e}} y_S,
\]
showing that $f_{p,\epsilon}$ is $\{c_0,\ldots,c_d\}$-quantized. Therefore, \Cref{lem:L0-L2} shows that $f_{p,\epsilon}$ is $O(\epsilon)$-close to Boolean.

We can now show that \Cref{itm:junta-constant} cannot be improved.
Notice that
\[
 \Pr[f_{p,\epsilon} = P(0)] \geq (1-p)^N \geq 4^{-\epsilon^{1/m_d}} = \Omega(1),
\]
using the fact that $(1-p)^{1/p}$ is monotone decreasing. Similarly,
\[
 \Pr[f_{p,\epsilon} = P(1)] \geq Np (1-p)^{N-1} = \Omega(\epsilon^{1/m_d}).
\]
It follows that for any constant $c$,
\[
 \EE[(f_{p,\epsilon} - c)^2] \geq \frac{1}{4} \Pr[f_{p,\epsilon} \in \{0,1\} \text{ and } f_{p,\epsilon} \neq \round(c, \{0,1\})] = \Omega(\epsilon^{1/m_d}).
\]
Choosing a sequence of $p,\epsilon$ with $\epsilon \to 0$ and $p \leq \epsilon^{1/m_d}$, this shows that the constant $m_d$ cannot be improved in \Cref{itm:junta-constant}.

\end{proof}

\subsection{Direct part}
\label{sec:junta-direct}

Let $f\colon (\{0,1\}^n, \mu_p) \to \{0,1\}$ be $\epsilon$-close to degree $d$, where $p \leq 1/2$. Applying \Cref{thm:structure}, we obtain a function $g\colon (\{0,1\}^n, \mu_p) \to \mathbb{R}$ which is $O(\epsilon)$-close to $f$ and satisfies the following properties:
\begin{enumerate}[(i)]
\item $g$ is a $(d,O(1/p))$-sparse junta.
\label{itm:g-sparse-junta}
\item $g(y) \in \{0,1\}$ whenever $y$ as weight at most $d$, and consequently $g$ is $B$-quantized for some finite set $B$ depending only on $d$.
\label{itm:g-quantized}
\item The number of minimal non-Boolean inputs of $g$ of weight $e$ is $O(\epsilon/p^e)$, for all $e$.
\label{itm:g-minimal-non-boolean}
\end{enumerate}

\Cref{itm:g-quantized} implies that $\tilde{g}(\emptyset) = g(0) \in \{0,1\}$. We assume without loss of generality that $g(0) = 0$: otherwise, replace $f$, $g$, and the approximating function $h$ by their ``negations'' $1-f,1-g,1-h$.

Let $\mathcal{M}$ consist of all inclusion-minimal sets in $\supp(g)$, and let $\mathcal{M}_e = \{ S \in \mathcal{M} : |S| = e\}$. The main part of the proof is the following Ramsey-theoretic lemma.

\begin{lemma} \label{lem:junta-aux}
For every $e \leq d$, either (i) $|\mathcal{M}_e| = O(1)$ or (ii) there exists a degree~$d$ polynomial $P$ such that $P(0) = g(0)$, $P(1) \neq P(0)$, and for every $m \leq m_d$ such that $m \ge 1$, there are $s \leq em$ and a collection $\mathcal{F}$ of $\Omega(|\mathcal{M}_e|^m)$ sets such that each $S \in \mathcal{F}$ satisfies the following:
\begin{enumerate}[(a)]
\item $S$ has size $s$.
\item $S$ is the union of $m$ sets in $\mathcal{M}_e$.
\item $g(\bbone_S) = P(m)$. 
\end{enumerate}
\end{lemma}

Let us see how this lemma implies the theorem. We start with the following corollary.

\begin{corollary} \label{cor:junta-aux}
For every $e \leq d$,
\[
 |\mathcal{M}_e| = O(1 + \epsilon^{1/m_d}/p^e).
\]
\end{corollary}
\begin{proof}
Let $e \leq d$, and apply \Cref{lem:junta-aux}. If Case~(i) holds then we are done, so suppose that Case~(ii) holds, for some $P$.

By the definition of $m_d$, there exists $m \leq m_d$ such that $P(m) \notin \{0,1\}$. 
Let $s,\mathcal{F}$ be as promised by the \namecref{lem:junta-aux} for $m$.
We would like to deduce a lower bound on $\Pr[g \notin \{0,1\}]$ from the lower bound on $|\mathcal{F}|$, and for this we apply the reverse union bound, \Cref{lem:reverse-union-bound}.

\Cref{lem:junta-aux} guarantees that
$\mathcal{F} \subseteq \supp(g)^{\ssum m}$. \Cref{lem:sparsity-elementary} shows that $\mathcal{G} = \supp(g)^{\ssum m}$ is $(md, O(1/p))$-sparse, and so the reverse union bound shows that
\[
 \Pr[\mathcal{E}] = \Omega(|\mathcal{F}| p^s) = \Omega(|\mathcal{F}| p^{em}),
\]
where $\mathcal{E}$ is the event that $y_S = 1$ for some $S \in \mathcal{F}$ and $y_R = 0$ for all $R \in \mathcal{G}$ such that $R \not\subseteq S$. If $\mathcal{E}$ happens then $g(y) = g(\bbone_S) = P(m) \notin \{0,1\}$, and so
\[
 \Pr[\mathcal{E}] \leq \Pr[g \notin \{0,1\}] \leq \Pr[g \neq f] = O(\epsilon).
\]
Consequently $|\mathcal{F}| = O(\epsilon/p^{em})$. Since $|\mathcal{F}| = \Omega(|\mathcal{M}_e|^m)$, we conclude that $|\mathcal{M}_e| = O(\epsilon^{1/m}/p^e) = O(\epsilon^{1/m_d}/p^e)$.
\end{proof}

We can now finish the proof of \Cref{itm:junta-constant} of \Cref{thm:junta}.

\begin{proof}[Proof of Item \ref{itm:junta-constant}]
If $g(y) \neq 0$ then $y_S = 1$ for some $S \in \mathcal{M}$. Therefore \Cref{cor:junta-aux} shows that
\[
 \Pr[g \neq 0] \leq \sum_{e=1}^d p^e |\mathcal{M}_e| =
 \sum_{e=1}^d O(p^e + \epsilon^{1/m_d}) = O(p + \epsilon^{1/m_d}),
\]
and so $\Pr[f \neq 0] \leq \Pr[f \neq g] + \Pr[g \neq 0] = O(\epsilon^{1/m_d} + p)$.

Since $g$ is a $(d,O(1/p))$-sparse junta, \Cref{lem:L0-L2-norm} implies that $\EE[(g-0)^2] = O(\epsilon^{1/m_d} + p)$, and so $\EE[(f-0)^2] \leq 2\EE[(f-g)^2] + 2\EE[(g-0)^2] = O(\epsilon^{1/m_d} + p)$.
\end{proof}

The proof of \Cref{itm:junta-boolean-degree-d} will proceed by bounding $|\supp_e(g)|$, where $\supp_e(g)$ consists of all sets in $\supp(g)$ of size $e$.

\begin{lemma} \label{lem:junta-supp}
There is a constant $N_d$, depending only on $d$, such that for every $e \leq d$,
\[
 |\supp_e(g)| = O(1 + \epsilon^{1/N_d}/p^e).
\]
\end{lemma}
\begin{proof}
\Cref{thm:nisan-szegedy} shows that any degree~$d$ function depending on more than $M_d$ coordinates cannot be Boolean. The union of $N_d = \binom{M_d}{\leq d} + 1$ distinct sets from $\supp_e(g)$ cannot have size at most $M_d$, and so if $A_1,\ldots,A_N \in \supp(g)$ are distinct then $g|_{A_1 \cup \cdots \cup A_N}$ cannot be Boolean.

If $|\supp_e(g)| < N_d$ then we are done, so suppose that $|\supp_e(g)| \geq N_d$. Then there are $\Omega(|\supp_e(g)|^{N_d})$ subsets of $\supp_e(g)$ of size $N_d$.
Let $\mathcal{F}$ consist of all unions of $N_d$ distinct sets from $\supp_e(g)$. Any such has size at most $eN_d$ and so can be written as a union of $N_d$ sets in at most $2^{eN_d^2}$ ways. Consequently, $|\mathcal{F}| = \Omega(|\supp_e(g)|^{N_d})$.

By construction, $\mathcal{F} \subseteq \mathcal{G} = \supp_e(g)^{\ssum N_d}$. Since $g$ is a $(d,O(1/p))$-sparse junta, \Cref{lem:sparsity-elementary} shows that $\mathcal{G}$ is $(eN_d,O(1/p))$-sparse. Applying the reverse union bound (\Cref{lem:reverse-union-bound}) with $p := 2p$, this shows that
\[
 \Pr_{\mu_{2p}}[\mathcal{E}] = \Omega(|\mathcal{F}| (2p)^{eN_d}),
\]
where $\mathcal{E}$ is the event that for some $S \in \mathcal{F}$ we have $y_S = 1$ and $y_R = 0$ for all $R \in \mathcal{G}$ such that $R \not\subseteq S$. Suppose that this event happens for $S$. Denoting $y = \bbone_T$, we have $g|_T = g|_S$, and so $g|_T$ is not Boolean. Furthermore, since $g|_T$ depends on $|S| \leq eN_d$ coordinates, we have $\Pr_{\mu_{1/2}}[g|_T \notin \{0,1\}] \geq 2^{-eN_d}$. This shows that
\[
 \Pr_{\mu_p}[g \notin \{0,1\}] \geq \Omega(|\mathcal{F}| p^{eN_d}) = \Omega(|\supp_e(g)|^{N_d} p^{eN_d}).
\]
Since $\Pr[g \notin \{0,1\}] \leq \Pr[g \neq f] = O(\epsilon)$, we conclude that
\[
 |\supp_e(g)| = O(\epsilon^{1/N_d}/p^e). \qedhere
\]
\end{proof}

We can now prove \Cref{itm:junta-boolean-degree-d} of \Cref{thm:junta}.

\begin{proof}[Proof of Item \ref{itm:junta-boolean-degree-d}]
\Cref{lem:junta-supp} shows that for some constant $K_d$ depending only on $d$,
we have $|\supp_e(g)| \leq K_d(1 + \epsilon^{1/N_d}/p^e)$. Write
\[
 g = \underbrace{\sum_{e\colon \epsilon^{1/N_d} \leq p^e} g^{=e}}_{g_J} + \underbrace{\sum_{e\colon \epsilon^{1/N_d} > p^e} g^{=e}}_{h},
\]
where $g^{=e}$ is obtained from the monomial expansion of $g$ by retaining only monomials of degree~$e$.

By construction, $|\supp_e(g)| \leq 2K_d$ for all $e$ in the first sum, and so $g_J$ is an $R_d$-junta, where $R_d = 2(d+1)K_d$. Similarly, $|\supp_e(g)| \leq 2K_d\epsilon^{1/N_d}/p^e$ for all $e$ in the second  sum, and so
\[
 \Pr[g \neq g_J] = \Pr[h \neq 0] \leq \sum_{e\colon \epsilon^{1/N_d} > p^e} p^e \cdot O(\epsilon^{1/N_d}/p^e) = O(\epsilon^{1/N_d}).
\]
It follows that $\Pr[f \neq g_J] \leq \Pr[f \neq g] + \Pr[g \neq g_J] = O(\epsilon^{1/N_d})$.

If $g_J$ is Boolean then we are done, assuming $\mexp_d \geq N_d$. Otherwise, since $g_J$ is an $R_d$-junta, $\Pr[g_J \neq f] \geq \Pr[g_J \notin \{0,1\}] \geq p^{R_d}$. Since $\Pr[g_J \neq f] = O(\epsilon^{1/N_d})$, we conclude that $p = O(\epsilon^{1/R_dN_d})$. Applying \Cref{itm:junta-constant}, this shows that $f$ is $O(\epsilon^{1/\mexp_d})$-close to the constant zero function, where $\mexp_d = \max(m_d, R_d N_d)$.
\end{proof}

\subsection{Proof of \texorpdfstring{\Cref{lem:junta-aux}}{\ref{lem:junta-aux}}}
\label{sec:junta-aux}

In order to complete the proof of \Cref{thm:junta}, we prove \Cref{lem:junta-aux}.

If $A_1,\ldots,A_m \in \mathcal{M}_e$ then we can write $g(A_1 \cup \cdots \cup A_m)$ as the sum of all monomial coefficients supported by $A_1 \cup \cdots \cup A_m$. Taking as an example the case $m = 2$, this corresponds to taking $e_1$ elements $B_1$ from $A_1$ and $e_2$ elements $B_2$ from $A_2$, where $e_1 + e_2 \leq d$, and summing over all $\tilde{g}(B_1\cup B_2)$ obtained in this way. If $A_1$ and $A_2$ are not disjoint then we might be summing the same monomial coefficient several times, and so we would like $A_1,A_2$, and more generally $A_1,\ldots,A_m$, to be disjoint for ``many'' tuples $A_1,\ldots,A_m$. However, this need not be the case: for example, it might be that all $A \in \mathcal{M}_e$ contain some element $a$. In the first step of the proof, we extract such a ``core'' inductively.

\begin{lemma} \label{lem:sunflower}
Let $e,m$ be parameters. Let $\mathcal{S}$ be a collection of sets of size $e$. There exist constants $C_{e,m},p_{e,m} > 0$ such that the following holds.

Either (i) $|\mathcal{S}| \leq C_{e,m}$, or (ii) there is a set $I$ such that with probability at least $p_{e,m}$, if we choose $A_1,\ldots,A_m \in \mathcal{S}$ at random then $A_1,\ldots,A_m \supseteq I$ and $A_1 \setminus I, \ldots, A_m \setminus I$ are disjoint.
\end{lemma}
\begin{proof}
The proof is by induction on $e$. If $e = 0$ then case~(i) clearly holds for $C_{e,m} = 1$, so assume that $e \geq 1$.

Suppose first that if we choose $A_1,\ldots,A_m \in \mathcal{S}$ at random, the probability that $A_1,\ldots,A_m$ are disjoint is at least $1/2$. If this happens then case~(ii) holds for $I = \emptyset$ assuming $p_{e,m} \leq 1/2$.

Suppose next that if we choose $A_1,\ldots,A_m \in \mathcal{S}$ at random, the probability that $A_1,\ldots,A_m$ are not all disjoint is at least $1/2$. If we now choose $i,j \in [m]$ such that $i \neq j$ then $A_i \cap A_j \neq \emptyset$ with probability at least $1/\binom{m}{2}$.

We can find a fixed choice $i = i_0$ for which the above holds. Thus if we choose $A \in \mathcal{S}$ at random, $A \cap A_{i_0} \neq \emptyset$ with probability at least $1/2\binom{m}{2}$. Consequently, one of the elements $a \in A_{i_0}$ is contained in a $q_{e,m} = 1/2e\binom{m}{2}$ fraction of the sets in $\mathcal{S}$.

Let $\mathcal{T}$ consist of $A \setminus \{a\}$ for all $A \in \mathcal{S}$ containing $a$, and apply the induction hypothesis. Thus, either (i') $|\mathcal{T}| \leq C_{e-1,m}$, or (ii') there exists $J$ such that  if we choose $A_1 \setminus \{a\},\ldots,A_m \setminus \{a\}$ at random from $\mathcal{T}$ then with probability at least $p_{e-1,m}$, all of these sets contain $J$, and are disjoint otherwise.

If case~(i') holds then $|\mathcal{S}| \leq |\mathcal{T}|/q_{e,m} \leq C_{e-1,m}/q_{e,m}$, and so case~(i) holds for $C_{e,m} = C_{e-1,m}/q_{e,m}$. If case~(ii') holds then case~(ii) holds for $I = J \cup \{a\}$ and $p_{e,m} = \min(1/2, q_{e,m}^m p_{e-1,m})$.
\end{proof}

Let $R$ be a constant to be chosen later,
and apply \Cref{lem:sunflower} with $e$, $m = R$, and $\mathcal{S} = \mathcal{M}_e$. If $|\mathcal{M}_e| \leq C_{e,R}$ then case~(i) holds, so suppose that there exists $I$ such that if we choose $A_1,\ldots,A_R \in \mathcal{M}_e$ at random then with probability at least $p_{e,R}$, all of $A_1,\ldots,A_R$ contain $I$, and $A_1 \setminus I, \ldots, A_R \setminus I$ are disjoint. If $|I| = e$ then the probability that all of $A_1, \ldots, A_R$ contain $I$ is $1/|\mathcal{M}_e|^R$, and so $|\mathcal{M}_e| \leq (1/p_{e,R})^{1/R}$, and again case~(i) holds.
We will show that if $|I| < e$ then case~(ii) holds. In this case, since $I$ is strictly contained in some element of $\mathcal{M}_e$, the definition of $\mathcal{M}_e$ shows that $g(\bbone_I) = 0$.

Suppose that $A_1,\ldots,A_R$ satisfy the constraints above, and let $A'_1,\ldots,A'_m$ be any $m \geq 1$ distinct elements chosen from $A_1,\ldots,A_R$. Then
\[
 g(A'_1 \cup \cdots \cup A'_m) =
 \sum_{\ell = 0}^d
 \sum_{1 \leq i_1 < \cdots < i_\ell \leq m}
 \sum_{\substack{J \subseteq I \\ D_{i_1} \subseteq A'_{i_1} \setminus I, \ldots, D_{i_\ell} \subseteq A'_{i_\ell} \setminus I \\ D_{i_1},\ldots,D_{i_\ell} \neq \emptyset \\ |D_{i_1}| + \cdots + |D_{i_\ell}| + |J| \leq d}} \tilde{g}(D_{i_1} \cup \cdots \cup D_{i_\ell} \cup J).
\]

This suggests coloring all tuples of up to $d$ elements with the value of the inner sum. Each such sum is the sum of $O(1)$ monomial coefficients of $g$, which is $B_d$-quantized for some finite set $B_d$ depending only upon on $d$, and therefore there are $C_d$ colors, for some constant $C_d$ depending only for $d$. The hypergraph Ramsey theorem now implies that for an appropriate value of $R$ depending only on $d$, there is a subset $A' \subseteq \{A_1,\ldots,A_R\}$ of size $m_d$ such that all tuples of $\ell \leq d$ elements from $A'$ have the same color $\gamma_\ell(A_1,\ldots,A_R)$.

Let $\gamma_0,\ldots,\gamma_\ell$ be a random choice of colors.
If we choose $A_1,\ldots,A_R \in \mathcal{M}_e$ at random, then with probability at least $p_{e,R}$, all of them contain $I$, and $A_1 \setminus I,\ldots,A_R \setminus I$ are disjoint. If this happens, then choose $A' \subseteq \{A_1,\ldots,A_R\}$ of size $m_d$ at random. With probability at least $1/\binom{R}{m_d}$, there are $\gamma'_1,\ldots,\gamma'_\ell$ such that all tuples of $\ell \leq d$ elements from $A'$ have color $\gamma'_\ell$. Furthermore, $\gamma'_\ell = \gamma_\ell$ for all $\ell$ with probability $1/C_d^{d+1}$.

Therefore, there is a choice of $\gamma_0,\ldots,\gamma_\ell$ such that if we choose $A_1,\ldots,A_{m_d} \in \mathcal{M}_e$ at random then with probability at least $p_{e,R}/\binom{R}{m_d}C_d^{d+1}$, all tuples of $\ell \leq d$ elements from $A' = \{A_1,\ldots,A_{m_d}\}$ have color $\gamma_\ell$. In particular, for all $m \leq m_d$ such that $m \geq 1$,
\[
 g(A_1 \cup \cdots \cup A_m) = \sum_{\ell \leq d} \binom{m}{\ell} \gamma_\ell.
\]
We call such $A_1,\ldots,A_m$ \emph{good}. The number of good tuples is thus $\Omega(|\mathcal{M}_e|^m)$.

Define the polynomial $P$ by
\[
 P(w) = \sum_{\ell=0}^d \binom{w}{\ell} \gamma_\ell.
\]
This is a polynomial of degree~$d$ such that
$P(0) = g(\bbone_I) = 0$ and $P(1) = g(\bbone_{A_1}) \neq 0$, since $A_1 \in \mathcal{M}_e$. The union $A_1 \cup \cdots \cup A_m$ has size $s = |I| + m(e-|I|) \leq em$.

Let $\mathcal{F}$ be the collection of all sets of the form $A_1 \cup \cdots \cup A_m$ for good $A_1,\ldots,A_m$. For any $S \in \mathcal{F}$, if $A_1 \cup \cdots \cup A_m = S$ then each $A_i$ is a subset of $S$, and so at most $2^{sm} \leq 2^{em^2}$ good tuples correspond to the same $S$. Since there are $\Omega(|\mathcal{M}_e|^m)$ good tuples, it follows that $|\mathcal{F}| = \Omega(|\mathcal{M}_e|^m)$.

\subsection*{Acknowledgements} We thank the anonymous reviewers for their careful reading of the manuscript, and for significantly improving some of the proofs. The second author thanks Mathews Boban for helpful discussions.

\printbibliography

\end{document}